%% file: THRI2017.tex
\documentclass[format=acmsmall, review=false, screen=true]{acmart}

\usepackage{booktabs} 

\usepackage[ruled]{algorithm2e} 

\SetAlFnt{\small}
\SetAlCapFnt{\small}
\SetAlCapNameFnt{\small}
\SetAlCapHSkip{0pt}
\IncMargin{-\parindent}

\usepackage{siunitx}
\usepackage{subcaption} 

\acmJournal{THRI}
\acmVolume{9}
\acmNumber{4}
\acmArticle{39}
\acmYear{2017}
\acmMonth{3}
\copyrightyear{2009}

\setcopyright{acmlicensed}

\acmDOI{0000001.0000001}

\received{February 2007}
\received[revised]{March 2009}
\received[accepted]{June 2009}

\begin{document}
\title[Adapting a General Purpose Social Robot for Paediatric Rehabilitation]
      {Adapting a General Purpose Social Robot for Paediatric Rehabilitation 
      through In-situ Design}  

\author{Felip Mart\'i Carrillo}
\orcid{0000-0003-0610-5375}
\affiliation{%
  \institution{Swinburne University of Technology}
  \city{Melbourne}
  \country{Australia}}
\affiliation{%
  \institution{Data61, CSIRO}
  \country{Australia}}
\email{fmarti@swin.edu.au}

\author{Joanna Butchart}
\affiliation{%
  \institution{Royal Children's Hospital}
  \city{Melbourne}
  \country{Australia}
}
\affiliation{%
  \institution{Murdoch Children's Research Institute}
  \city{Melbourne}
  \country{Australia}
}
\email{jo.butchart@rch.org.au}

\author{Sarah Knight} 
\affiliation{%
  \institution{Murdoch Children's Research Institute}
  \city{Melbourne}
  \country{Australia}
}
\affiliation{%
  \institution{Royal Children's Hospital}
  \city{Melbourne}
  \country{Australia}
}
\email{sarah.knight@mcri.edu.au}

\author{Adam Scheinberg}
\affiliation{%
  \institution{Royal Children's Hospital}
  \city{Melbourne}
  \country{Australia}
}
\affiliation{%
  \institution{Murdoch Children's Research Institute}
  \city{Melbourne}
  \country{Australia}
}
\email{adam.scheinberg@rch.org.au}

\author{Lisa Wise}
\affiliation{%
  \institution{Swinburne University of Technology}
  \city{Melbourne}
  \country{Australia}}
\email{lwise@swin.edu.au}

\author{Leon Sterling}
\affiliation{%
  \institution{Swinburne University of Technology}
  \city{Melbourne}
  \country{Australia}}
\email{lsterling@swin.edu.au}

\author{Chris McCarthy}
\affiliation{%
  \institution{Swinburne University of Technology}
  \city{Melbourne}
  \country{Australia}}
\email{cdmccarthy@swin.edu.au}

\begin{abstract}
Socially Assistive Robots (SARs) offer great promise for improving outcomes 
in paediatric rehabilitation. 
However, the design of software and interactive capabilities for SARs must be
carefully considered in the context of their intended clinical use. 
While previous work has explored specific roles and functionalities to support paediatric rehabilitation, 
few have considered the design of such capabilities in the  
context of ongoing clinical deployment.
In this paper we present a two-phase \emph{In-situ design}
process for SARs in health care, emphasising stakeholder engagement and on-site development.
We explore this in the context of developing the humanoid social robot NAO  
as a socially assistive rehabilitation aid for children with cerebral palsy.
We present and evaluate our design process, outcomes achieved, and preliminary results from ongoing 
clinical testing with 9 patients and 5 therapists over 14 sessions.  We argue that our  
in-situ Design methodology has been central to the rapid and successful deployment of our system.
\end{abstract}

%
%

 \ccsdesc[500]{Human-centered computing~Field studies}
 \ccsdesc[500]{Computer systems organization~Robotics}
 \ccsdesc[300]{Applied computing~Health care information systems}
 \ccsdesc[100]{Social and professional topics~Children}

%
%

\keywords{
    In-Situ Design; Socially Assistive Robots; Rehabilitation; Health Care
}

\maketitle

\renewcommand{\shortauthors}{F. Mart\'i Carrillo et al.}

\input{THRI2017-body}

\end{document}

%% file: THRI2017-body.tex
\section{Introduction}
\label{sec:introduction}

Rehabilitation outcomes rely critically on patients adhering to a prescribed 
set of rehabilitation exercises~\cite{thomason2013rehabilitation}. 
When those patients are children, maintaining compliance and focus while performing 
what can often be tiring, uncomfortable and repetitive exercise programs presents
a significant challenge~\cite{Rainae2005, Plant2007Predictors}.   
While therapists and carers are well equipped with skills and experience to maintain a child's motivation, 
this takes considerable time and resources \cite{McDonald1996ImpactMajorEvents}.  
Therapists are not always able to attend each prescribed exercise session, and even  when present,
results are not always positive.  

Socially Assistive Robots (SARs) are increasingly being considered to support
a range of  health care delivery needs.  
SARs provide assistance primarily through social interaction and engagement~\cite{feil2005defining},
 i.e children suffering form serious illness~\cite{Bers1998StoretellingCardiac}. 
SARs have shown promising results for improving mood, reducing stress, and encouraging 
communication for children on the autism spectrum~\cite{robins2005robotic}, 
in rehabilitation~\cite{calderita2013therapist,mejias2013ursus}, 
for encouraging exercise in older adults~\cite{fasola2013socially},
and in post-stroke rehabilitation~\cite{wade2011using}. 

Paediatric rehabilitation presents an ideal context for the application of SARs. 
Previous work suggests SARs may provide therapeutic benefits for patients through 
increased focus and compliance~\cite{fridin2014robotics, Kozyavkin2014}.
However, no formal clinical evaluation of the therapeutic benefits of SAR's for rehabilitation currently 
exists. This requires development beyond proof-of-concept, 
with clear clinical use-cases identified.  While previous work has
explored specific roles and functionalities to support paediatric rehabilitation (e.g., \cite{mejias2013ursus,calderita2013therapist,malik2015human,borovac2016human}) 
few have considered the design of such capabilities in the context of ongoing clinical deployment.
Addressing this gap is critical to understanding the clinical context SARs must operate in, 
and for establishing the long term legitimacy of SARs as 
effective and usable therapeutic aids with therapists and care-givers.

We are developing software to adapt the humanoid robot NAO
as  a therapeutic aid for paediatric rehabilitation, and evaluating its effectiveness.   
In partnership with a busy paediatric rehabilitation clinic of The Royal Children's Hospital, Melbourne, Australia, 
we are developing a range of interactive and demonstrative behaviours 
for NAO to enhance patient compliance, motivation and emotional well-being during therapy sessions.   
We aim to deploy NAO robots as both a therapist's assistant during sessions, and as a proxy to therapists when they are unable to attend 
(eg., on-ward after-hours, or at home).  
The near-term goal is thus to autonomously support independent exercise programs on the ward, 
before then extending the system's use to supporting prescribed rehabilitation programs at home.
To this end, we are determining roles and developing robust interactive capabilities that allow NAO to guide 
patients through complete exercise sessions without engineer monitoring 
(or \emph{Wizard-of-Oz} control), or additional hardware (e.g., external sensors).

In this paper we report on 23 months of progress designing and developing software for NAO as 
a therapeutic aid for paediatric rehabilitation. 
Focused on the needs of large scale clinical deployment,
we outline key requirements for an SAR operating as a stand-alone therapeutic aid for ongoing use in a clinical setting.
We present a two-phase in-situ design process, including both exploration of roles 
and requirements, from which a base-level stand-alone prototype system has been derived. 
To our knowledge, this is the first design of an SAR 
for rehabilitation that explicitly incorporates patients, carers and therapists in the design process,
and is focussed on the design of roles and capabilities for ongoing use
in a clinical setting. 
Our prototype system is now 
deployed  in weekly therapy sessions, leading predominantly patients with cerebral palsy through prescribed exercise programs 
of up to 30 minutes without engineer intervention.

The paper is structured as follows. Section \ref{sec:background} gives background and
an overview of previous work.  Section \ref{sec:methodology} outlines our in-situ 
design methodology, listing derived roles and requirements for the system from Phase 1 of this process.
Section \ref{sec:technical} and \ref{sec:designdecisions} 
provides a technical overview of the current system deployed in 
Phase 2 development, and key design choices and considerations.  We present our clinical testing
setup, a discussion of preliminary Phase 2 results and feedback in Section \ref{sec:clinicalsessions}.  
Our conclusions are presented in Section \ref{sec:Conclusions}.

\section{Background and Related Work}
\label{sec:background}

\subsection{Socially Assistive Robots in Paediatric Rehabilitation}

A number of groups have considered Socially Assistive Robots (SARs) for rehabilitation, focussing
primarily on technical developments and evaluations of  proof-of-concept systems.  
``Ursus'' \cite{mejias2013ursus} is a combination of  
a low-cost robot and an augmented reality device to assist upper limb rehabilitation exercises
for children with cerebral palsy (CP) .  The system was evaluated in single sessions with six patients, with feedback  suggesting the SAR 
enhanced enjoyment, and had a positive impact on rehabilitation sessions.
``Therapist''~\cite{calderita2013therapist}, the evolution of the ``Ursus'' robot platform, 
provides a virtual reality video game,  and  exercise demonstrations for upper-limb exercises.   A thorough evaluation of the
 system's  cognitive framework  (eg. speech/emotion recognition, human detection, etc.) is provided from both
lab-based and in-the-field experiments.  
 Exercise demonstration and robot mirroring is also proposed by Fridin et al.  \cite{fridin2014robotics} to assist groups of paediatric patients, 
 and Malik et al.~\cite{malik2015human} who implement three different exercise demonstrations  (Sit to Stand, Balancing, and Ball kicking).
 ``MARKO'' ~\cite{borovac2016human},  a  
robot sitting on a horse-like mobile platform, 
is designed to assist rehabilitation for patients with CP  in gross motor skill exercises, fine motor skills and speech exercises.

While previous systems have been tested with patients, no  existing SAR has been  deployed  as part of the ongoing
rehabilitation program of paediatric patients. 

\subsection{In-situ design and evaluation in the wild}
Human-robot and Human-computer interaction researchers have previously reported issues in the
extrapolation of lab-based evaluations into real world contexts.   In the hospital context, Multu and Forlizzi \cite{Mutlu2008ROR}
describe the rejection of deployed autonomous delivery robots by hospital staff due to interruptions and distractions
inflicted on them when performing higher priority tasks.  Such issues have promoted the use of  \emph{in-situ} design and
\emph{in the wild} evaluation methodologies in which new technologies are designed and evaluated in-place and under the conditions 
of their intended use~\cite{Rogers2011InTheWild}.

Museums and public spaces have been a popular target of in-the-wild HRI design and evaluation \cite{burgard1999experiences,thrun2000probabilistic},
as well as in the home.  Kidd and Brazel \cite{kidd2008robots} report  on the in-situ design and evaluation of
a weight loss coach robot, benchmarking it against a stand-alone computer, and a traditional paper log.  
They show a two-fold increase in exercise time for participants using the robot, compared to those using the aids.
H\"uttenrauch  et al. \cite{huttenrauch2009art}  study  participant interaction patterns with a mobile robot in a home guided tour. 
More recently, Pripfl et al. \cite{pripfl2016results} report on the results of an in-the-wild evaluation of a service robot deployed in the homes of 18 elderly participants.  
Their findings highlighted issues with both technical performance of the system, and participant perceptions of the robot as a toy rather than an aid.

{\u S}abanovi{\'c} et al. \cite{vsabanovic2014designing} report on the in-situ design and development of a robot to manage break times in an office environment.
They note benefits for identifying contextual issues impacting robot use, and for including users in the design process even when evaluating with incomplete and non-robust
prototypes.  CERO \cite{1045615} was used to assist in the transport of objects in an office environment for partially motion-impaired users over a 3 month study.
The in-situ evaluation of the prototype identified important factors not considered previously such as physical space 
limitations and bystander engagement. 
An in-situ HRI study by Michalowski et al. \cite{Michalowski2007}  examined  social engagement with two social robots in a conference setting.  This evaluation 
identified flawed  design assumptions, leading to new ideas and improvements in the robot's interaction effectiveness.

The in-situ design of SARs in health settings is less common, though examples of evaluation during deployment exist. 
Studies using the  seal robot PARO,  for example, have shown benefits for improving mood, reducing stress and encouraging social engagement for residents in an aged-care facility.
Such studies have performed evaluations over  5 weeks \cite{wada2004effects}; 4 months \cite{kidd2006sociable}; and 1 year \cite{wada2005psychological}.
In-situ studies have also evaluated PARO  as a therapeutic aid for people with dementia \cite{giusti2008robots, moyle2013exploring, chang2013situated}.
Most closely aligned to our application, Plaisant et al. \cite{Plaisant2000} employed a Participatory Design approach 
in the design of an SAR prototype to enhance rehabilitation outcomes with children.
They iteratively evaluated their prototypes during the design sessions with their intended final users.
However, unlike our approach, they did not deploy the SAR to lead sessions, or as part of the ongoing care delivery.

Our contributions differ from previous work in the following distinct ways.  Firstly, we focus specifically on  the design of an SAR for ongoing therapeutic use
by a therapist or care-giver, and for leading entire therapy sessions with children.   Moreover, 
we adapt and evaluate   a general purpose social robot (NAO)  as a stand-alone system, outlining  design decisions and requirement compromises to achieve this. 
Finally, we outline and evaluate our design process for SARs in rehabilitation,
noting specific design outcomes resulting from our in-situ design and evaluation, and the explicit inclusion of stakeholders in this design process.

\section{Design Process}
\label{sec:methodology}
We have engaged in a two-phase in-situ design process,  incorporating
both exploratory and iterative prototyping, and frequent engagement with key stakeholders.  Below we describe
the project context, stakeholders and the implementation of these two phases of development.

\subsection{Project Setting}
The proposed Socially Assistive Robot (SAR) system is being developed in close partnership with a busy paediatric 
rehabilitation clinic in a city-based children's hospital. The rehabilitation clinic 
consists of 25 full-time equivalent clinical staff servicing, on average, 
180 inpatients annually, as well as several thousand outpatient sessions.   
Patients seen at the clinic range from those recovering from physical injury and illness 
to those being treated for specific chronic disabilities.  Inpatients generally undergo intensive rehabilitation 
programs requiring multiple sessions of rehabilitation per day.   
While some sessions are supervised by physiotherapist staff, others may be facilitated by on-ward nursing staff, 
or the patient's parent.  A particularly prominent patient group are those children with cerebral palsy (CP).  
In many cases, orthopaedic surgery is required to correct secondary musculoskeletal problems which impact on gait and function.   
Such patients typically undergo up to three rehabilitation sessions per day, over a 2 to 3 week period \cite{thomason2013single}.

\subsection{Stakeholders}
We identified the following four groups as key stakeholders in the development
of the SAR for rehabilitation.  
\begin{description}
\item [Patients:] the primary beneficiaries of the SAR through potentially increased motivation and sustained emotional well-being, faster recovery time and improved rehabilitation outcomes.   
They are chief determinants of the SAR's interaction design. 

\item [Therapists/Healthcare providers:] primary users of the system,
with use-cases spanning both in-session use as well as pre-configuration
for sessions without their direct supervision.   They are determinants of
the SAR's therapeutic assistance, correctness (eg., exercise demonstrations),
 usability, integration and fitness for purpose.

\item [Parents/Guardians:] holders of primary duty of care for patients, are often present during therapy sessions 
and tasked with ensuring rehabilitation exercises are performed outside of formal therapy sessions (e.g., on-ward, after-hours).    
They are thus targeted end-users of the system, and determinants of the system's usability, and fitness for purpose.

\item [Technology Developers:] engage with all other stakeholders to 
determine the SAR system requirements, design and implement interactive 
behaviours and operate the SAR during development and testing.  They gather
feedback from other stakeholders, assess the system's technical
performance, and the feasibility of identified roles and requirements.
\end{description}

\subsection{Design and Development}
\begin{figure}[htb]
\centering
\includegraphics[trim=0 350 0 0,clip,width=0.95\columnwidth]{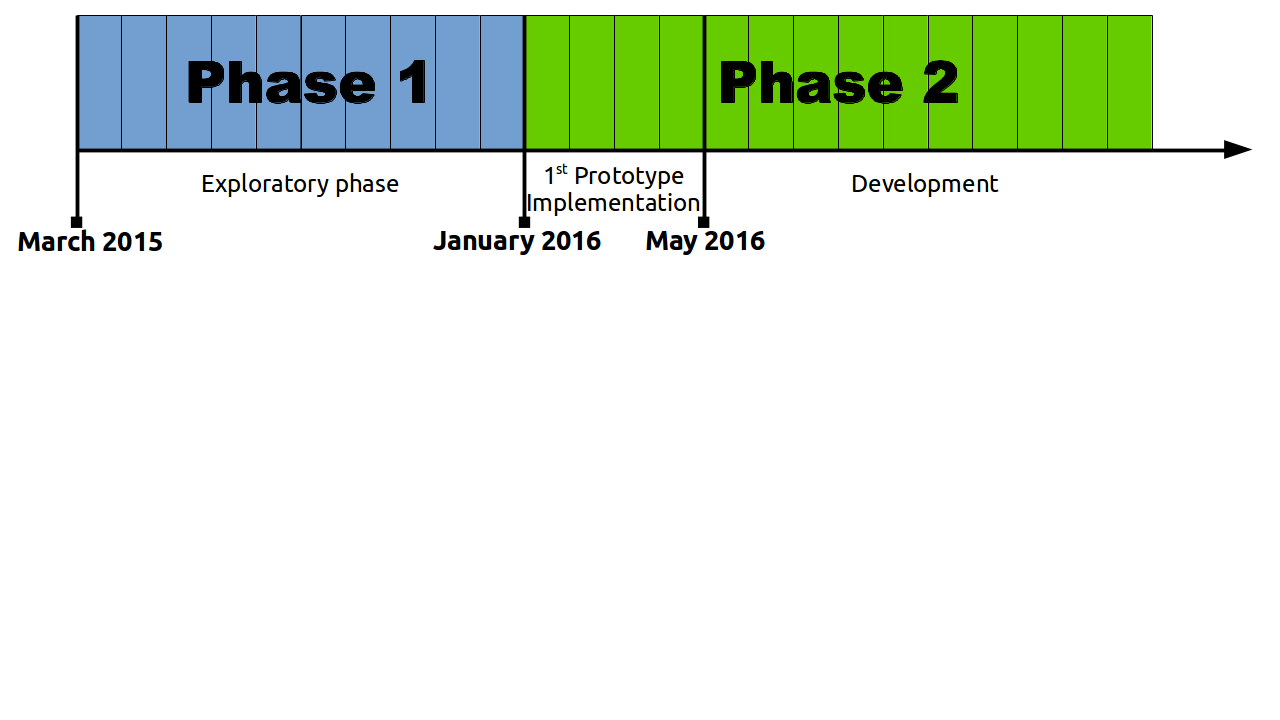}  
\caption{Timeline of the project until the current stage.}
\label{fig:Phases}
\end{figure}

Our design approach has consisted of two phases. The first, an exploratory phase to elicit basic requirements, ran for 10 months between March 2015 and January 2016.
The second phase, involving the  iterative  development and in-situ evaluation of a first prototype implementation, began in March 2016, and is ongoing.  Through these design phases, a prototype for formal clinical trials  is being targeted. 
Figure~\ref{fig:Phases} shows the timeline of development to date. We describe both phases below.

\subsection{Phase 1: Exploration}
The initial phase of the SAR's design, previously described in  \cite{mccarthy2015robots}, prioritised two key activities: regular and frequent (weekly) stakeholder engagement, 
and rapid prototyping and mock-ups (via \emph{Wizard-of-Oz} control) of proposed roles and capabilities.  
Both activities were conducted primarily on-site, in the context of the SAR's intended deployment.   

A regular weekly pattern of visits to the clinic was established in the early weeks of the phase.  
Each Tuesday morning attending research team members (typically two) setup NAO 
in a publicly visible and accessible location, close to consultation rooms with high visibility to patients, 
their families, and therapists.  This facilitated regular, albeit brief, discussions with therapists and parents at the beginning.  
Patient interactions were initially also brief, unstructured and intermittent, typically occurring during their time waiting for a consultation with therapists.
The  use of \emph{Wizard-of-Oz} control via a laptop with wireless link to NAO, 
allowed the SAR to meet the immediate needs of particular interactions.

Early engagement suggested how to overcome the technology limits and foster effective engagement with patients. 
It facilitated development of core exercise demonstrations. Therapists were actively engaged in this
process, initially through requests to critique NAO's execution of exercises, 
and also invited to physically manipulate the robot's limbs to both correct 
and explore the physical capabilities and limitations of the system.  

In the second half of the phase, therapist  engagement evolved into a cycle of iterative development 
in which a therapist directly programmed specific exercises by positioning the robot into key poses, 
from which robot joint positions were immediately recorded and time sequenced. 
New exercises were rapidly developed via this process on-site, with refinements made between clinic visits.
During this second half of Phase 1, observations determined  specific roles (outlined in Section~\ref{subsec:roles}) 
based on the robot's capabilities, and the derivation of requirements for an SAR (Section~\ref{subsec:requirements}) for ongoing clinical use.

Patient engagement also progressed from non-specific patient interactions driven primarily by general interest and the novelty of the robot in the waiting area, 
to the active inclusion of NAO in therapist-selected patient sessions.
Pre-built exercise demonstrations were sequenced in accordance with therapist specifications, 
and trialled in sessions with technical support.  
Early scripting of SAR behaviours was done using the vendor-supplied graphical development environment, Choreographe. 
This visual programming environment, while limiting in some technical respects due to its highly abstracted \emph{block-style}
 programming,
allowed different Technical Developers to interchangeably operate NAO 
without requiring specialised knowledge of  underlying system complexities, 
thereby increasing the pool of developers who could assist in this exploratory phase.
This supported the maintaining of regular weekly visits throughout Phase 1, and diversified 
interactions between developers and all non-technical stakeholders.

\subsection{Phase 2: Development}
Phase 2 is ongoing,  prioritising the in-situ iterative development and evaluation of a stand-alone prototype in preparation
for formal clinical evaluation. 
As such, focus has been placed on the realisation of a minimum viable SAR based on the roles determined in Phase 1,
and the identified key requirements in both phases for an SAR in rehabilitation \cite{marti2017insitu}.

Regular weekly patient sessions with
NAO have been scheduled in which \emph{Wizard-of-Oz} control and engineering support has been removed from the SAR's operation,
thus focusing on the needs of ongoing stand-alone operation in a clinical setting. 
Phase 2 aims to develop the system to be under the sole operation of therapists, parents and/or other care-givers. 

Phase 1 established cerebral palsy as a well suited initial target for clinical evaluation.   
Phase 2 has thus  focussed  on a system 
capable of leading sessions for patients with cerebral palsy undergoing 
post-operative rehabilitation.  Exercise capabilities predominantly target lower-limb strengthening in accordance with the typical prescribed program of rehabilitation
for this patient group.

Patients, therapists and parents not involved in Phase 1 have been formally recruited and consented 
to participate in this phase of the study. Data is gathered via questionnaires with all 
stakeholders at the completion of each session, along with observation notes recorded during each session 
(detailed in Section~\ref{subsec:Data}).
Attending researchers have observed from an adjacent room with one-way mirror.  
We discuss the details of clinical sessions in Section \ref{sec:clinicalsessions}.

\subsection{\label{subsec:roles}Derived Roles}
Therapist consultation and observation during Phase 1 determined 
four specific roles encompassing the base-level capabilities the SAR must 
provide to serve as an effective therapeutic aid in rehabilitation sessions. 

\begin{description}
\item[Demonstrator:] At the beginning of each exercise set, the SAR performs the exercise in front of the child. 
The SAR also provides verbal instructions to emphasise important aspects of the exercise.

\item[Motivator:]  The SAR provides verbal encouragement at the beginning of each session, 
as well as before and during each prescribed exercise.  
Enticements such as entertainment through music, dancing and joke telling are also 
offered upon completion of exercise sets.  

\item[Companion:] The SAR delivers personalised introductory statements at the beginning of the session 
to build rapport and establish itself as a joint participant in the session.   
As the child performs each exercise set, the SAR joins in and delivers empathetic and encouraging statements 
acknowledging the child's progress.

\item[Coach:] The SAR guides the patient through the prescribed session by scheduling and coordinating the execution of the above roles to deliver a complete session of therapy.  
The system paces the delivery in accordance with the patient and therapist/carer responses.
\end{description}

\subsection{Derived Requirements}
\label{subsec:requirements}
To support the above roles, Phase 1 identified the following system requirements. 
\subsubsection{Configurability:} 
Therapists and Technology Developers in Phase 1 both identified the need for configurability of the system to realise 
a stand-alone SAR for rehabilitation.   Early feedback from therapists requested a system based on current practise in which session schedules are produced
by selecting activities from  a list.   
Configuration thus needs to allow pre-selection of exercises to perform,  the number of repetitions, speed of execution, entertainment modules,
as well as personalisation of the session with the patient.

\subsubsection{Stability:}
Therapists and Technology Developers jointly determined that exercise demonstrations and 
general SAR actions must operate with a high degree of certainty in order to minimise session interruption 
and distraction.  In the context of an off-the-shelf general purpose social robot,  physical characteristics impacting this are not modifiable,
and thus must be carefully managed within the programmed movements of the system.

\subsubsection{Adaptability:}
To ensure  therapeutic assistance is aligned with the patient's needs,
the SAR should be adaptable to the presenting condition of the patient during care delivery.  
It was observed in Phase 1 that therapists prescribe exercises before a session, but assess and adjust activities during the session.  
Therapists noted that an effective SAR for rehabilitation should provide mechanisms for dynamic adjustment of activity settings, 
including number of repetitions, speed and sequence order. Verbal instructions must adjust accordingly.

\subsubsection{Interaction:}
Observations in Phase 1 indicated a general desire of patients to interact with the robot,
and this should be facilitated often.  Basic interaction with the SAR should always be supported for therapists/carers and patients throughout the session.
Challenges observed with speech recognition during Phase 1 made clear that interaction should be multimodal (eg. verbal, tactile, etc.) to cater for varying patient needs.
This will support Adaptability, Responsiveness and maintain  patient engagement.

\subsubsection{Integration:} 
Previous work (eg., Mutlu and Forlizzi \cite{Mutlu2008ROR}) and Phase 1 observations highlighted the need to ensure setup and use of the SAR was well integrated with existing clinical
practise, and the general operating conditions of a busy hospital-based rehabilitation clinic.   Therapists and Technology Developers together determined that
the SAR must be easily setup by therapists and care-givers, be portable and transportable by a single person, and operable by carers with minimal training requirements.

\subsubsection{Responsiveness:} 
Observations by Technology Developers in Phase 1 and early Phase 2 sessions  indicated that  a lack of responsiveness to unprompted verbal statements from patients 
may diminish the perceived authenticity of the SAR's role as a companion.
Observations also highlighted that the implementation of responses should also incorporate  awareness of the patient's mood and progress to 
support the SAR's motivator role.

\subsubsection{Stand-alone:} 
Therapists and Technology Developers jointly agreed  that the system should be operable without 
engineering support, \emph{Wizard-of-Oz} control, or additional hardware to meet the needs of flexible and un-hindered ongoing use. 
SAR activities requiring human assistance should also be minimised to ensure carer focus remains primarily on the patient.
Therapists also expressed a strong desire to have the SAR present and ready to use at the hospital at all times.

\subsubsection{Robustness and Endurance:}
To meet the needs of leading a rehabilitation sessions, therapists and technology developers  determined the system needs
to operate continuously and  for a minimum of 30 minutes without engineer intervention. 
To support the stand-alone requirement, unforeseen interruptions such as falls, slippage, or unintended/incorrect user interactions should also be recoverable from, 
either automatically, or through a clearly understood set of instructions for the therapist and/or care-giver to follow.
\\

In Sections \ref{sec:technical} and \ref{sec:designdecisions} we outline the technical implementation and key design decisions 
to maximise the realisation of these baseline roles and requirements.

\section{SAR Prototype Implementation}
\label{sec:technical}

\subsection{Software Modules}

Our prototype software for the NAO robot platform utilises the Robot Operating System (ROS), 
an open-source robotics framework.  ROS was chosen on the basis of its extensibility and 
strong support for simplified communication between different tools, 
and devices in a robotic system~\cite{quigley2009ros}.

\begin{figure}[htb]
\centering
\includegraphics[trim=0 550 825 0,clip,width=0.95\columnwidth]{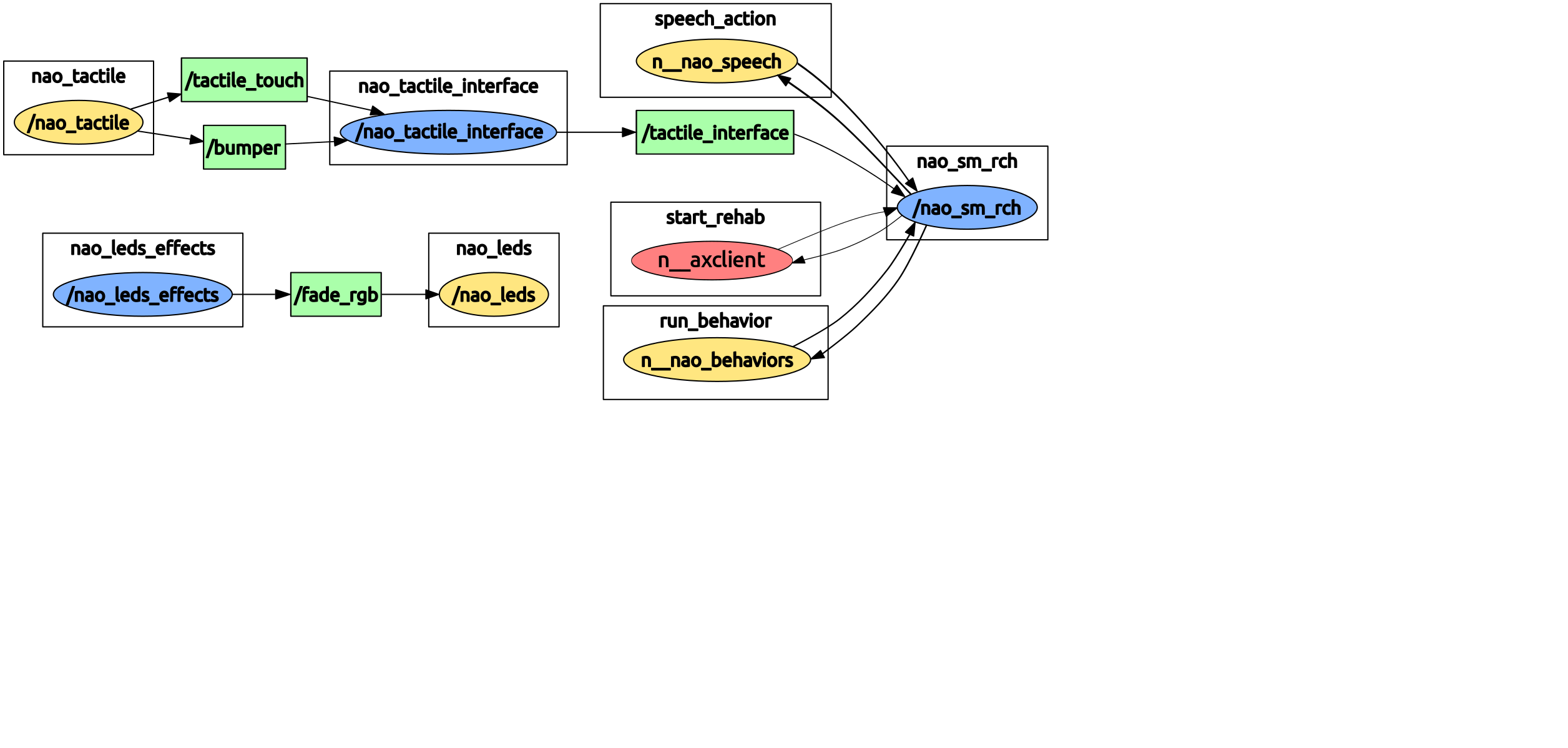}  
\caption{ROS modules simplified for the SAR. In blue:  ROS nodes implemented for the prototype.
In yellow: ROS nodes that connect ROS with NAO's server . 
In green:  the ROS communication topics.
In red: the graphical actionlib client to start the session from a remote computer.}
\label{fig:ROSModules}
\end{figure}

Figure~\ref{fig:ROSModules} shows some of the basic modules of the NAO robot for ROS and the  
three modules implemented in our system. We briefly describe each below.

\subsubsection{nao\_sm\_rch}
 is the main module of the system, incorporating all rehabilitation 
activity scenarios, including speech, lower body exercises, games, and dances. 
We implement this as a finite state machine initiating specific scenarios 
via connections to other nodes of the system.

Services such as \texttt{run\_behavior} or \texttt{speech\_action} are called from this node
in order to execute a predefined movement, or to make the robot speak.
To assist data collection, the module also maintains a logfile 
tracking all the exercises executed, timing data, and user-inputs.

\subsubsection{nao\_tactile\_interface}
 is implemented as a ROS service to capture and detect inputs to the system such as from touch 
sensors and bumpers using the \texttt{nao\_tactile} library.  This interface detects single, 
double and long button clicks, allowing numerous different responses to be invoked.

\subsubsection{nao\_leds\_effects}
provides visual prompts and conveys the system state.
We have configured this service using the ROS NAO library \texttt{nao\_leds} with 5 different 
LED effects that are activated to cue the need for the robot's head to be tapped in order 
to continue the session, or to indicate a session configuration file is being loaded.

\subsubsection{Other nodes}
Figure~\ref{fig:ROSModules} shows other ROS libraries that we are using such as \texttt{nao\_leds},
\texttt{nao\_tactile}, \texttt{run\_behavior} and \texttt{speech\_action}.
The robot is configured and started using the \texttt{start\_rehab} action library.

\subsection{Activity Scenarios}
\label{sec:ActiScen}
Our current prototype for Phase 2 trials implements 16 different activity scenarios 
to support the roles outlined in Section~\ref{subsec:roles}.
Activity scenarios are all the rehabilitation exercises (N=13), 
plus an introductory speech delivery, a toy relay game, and entertainment routines.
In the introductory speech the robot introduces itself to the patient, 
or greets a patient it has previously interacted with.  
In addition to statements explaining what is planned for the session, 
the scenario includes jokes and pre-programmed dialogue  to foster rapport building.
Several introductory speeches can be selected from to reduce repetition over multiple sessions.

Sessions consist of multiple exercises, each involving several sets and 
repetitions. 
Adjustments to exercise speed,
if requested during the session, can be changed by the carer using the \emph{Tactile Interface},
explained in more detail in Section~\ref{subsec:TacInt}.
For each exercise, the SAR presents a demonstration while
explaining key features of the exercise.  The patient is then invited to join
the SAR in completing a set together.  During exercise execution the SAR provides
encouraging and therapist-selected reminders about key aspects of each exercise 
(Section~\ref{subsec:RobSpeech}).
At the completion of each set, the SAR requests the patient (or carer)
tap its head to continue.  
The SAR asks for help when human assistance is required to setup a particular
activity (Section~\ref{subsec:Help}). 

\begin{figure}[htb]
\centering
\includegraphics[trim=0 0 0 0,clip,width=0.7\columnwidth]{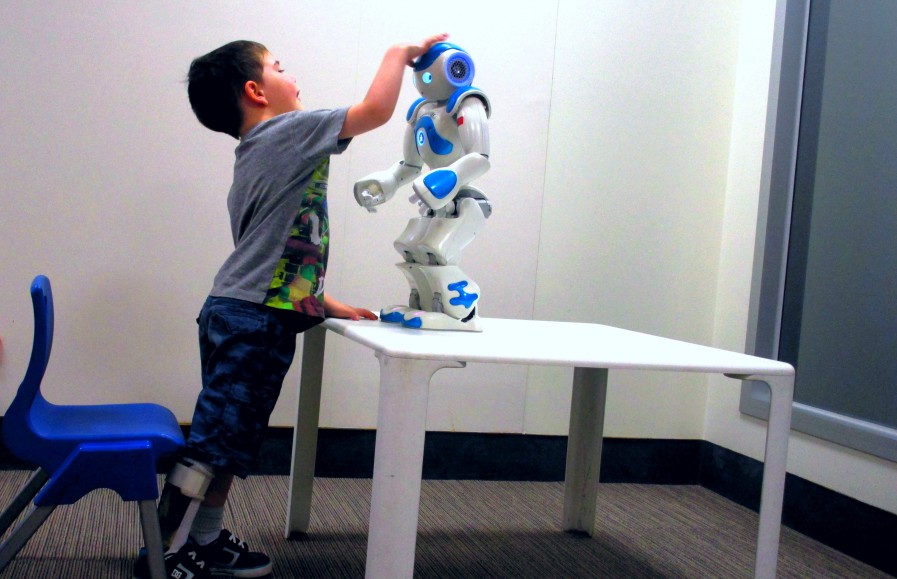}  
\caption{\textbf{Sit-to-Stand} exercise: The patient taps the robot's head to initiate the robot's 
stand up sit down actions while the child follows.
[Guardian consent provided]}
\label{fig:SitStands}
\end{figure}

\begin{figure}[htbp]
    \begin{subfigure}[b]{0.3\textwidth}
        \subcaption{\textbf{Bridge}}
        \includegraphics[width=\columnwidth]{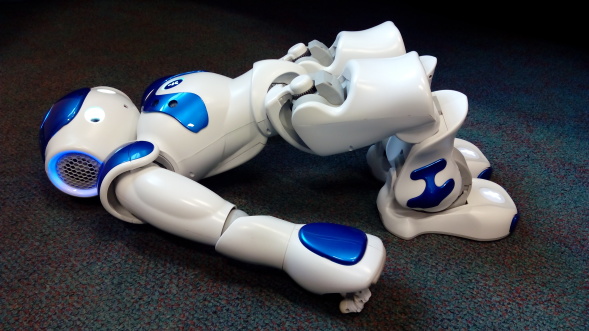} 
    \end{subfigure}
    \begin{subfigure}[b]{0.3\textwidth}
        \subcaption{\textbf{Hip Abduction Laying}}
        \includegraphics[width=\columnwidth]{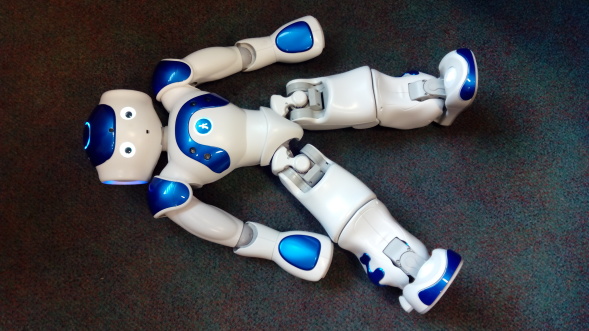} 
    \end{subfigure}
    \begin{subfigure}[b]{0.3\textwidth}
        \subcaption{\textbf{Hip Abduction on Side}}
        \includegraphics[width=\columnwidth]{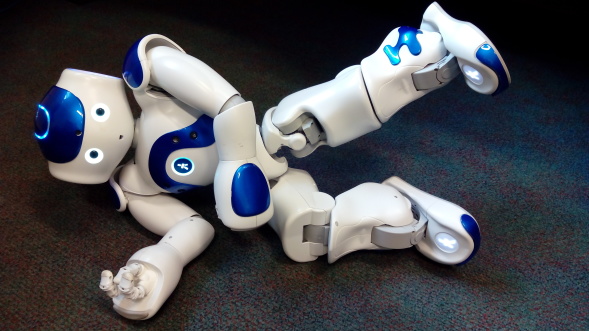}  
    \end{subfigure}

    \begin{subfigure}[b]{0.3\textwidth}
        \subcaption{\textbf{Single Bridge}}
        \includegraphics[width=\columnwidth]{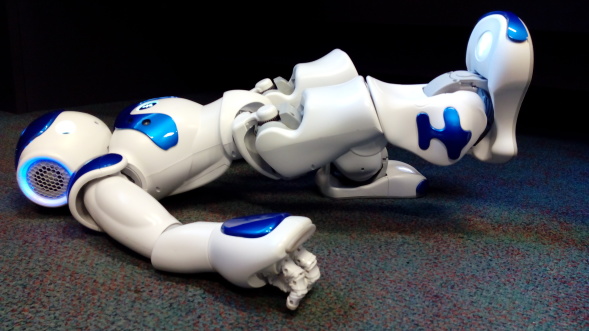} 
    \end{subfigure}
    \begin{subfigure}[b]{0.3\textwidth}
        \subcaption{\textbf{Hip Extension Easy}}
        \includegraphics[width=\columnwidth]{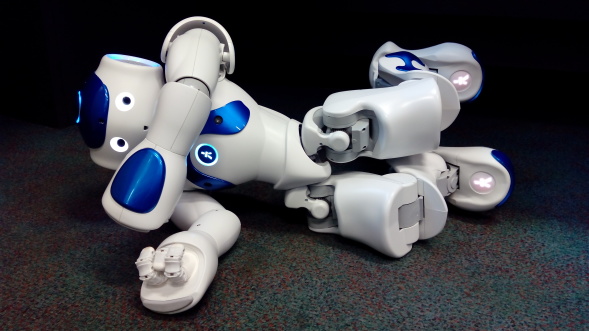} 
    \end{subfigure}
    \begin{subfigure}[b]{0.3\textwidth}
        \subcaption{\textbf{Hip Extension Hard}}
        \includegraphics[width=\columnwidth]{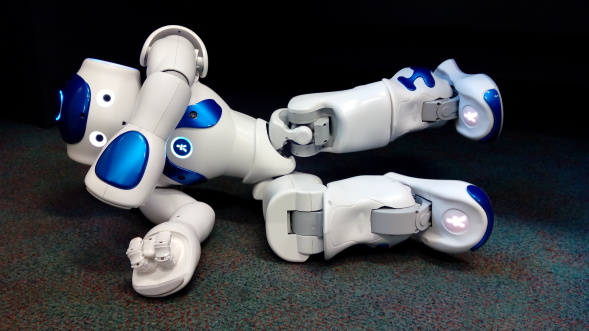}  
    \end{subfigure}

    \begin{subfigure}[b]{0.3\textwidth}
        \subcaption{\textbf{Hip Knee Flexion Sliding}}
        \includegraphics[width=\columnwidth]{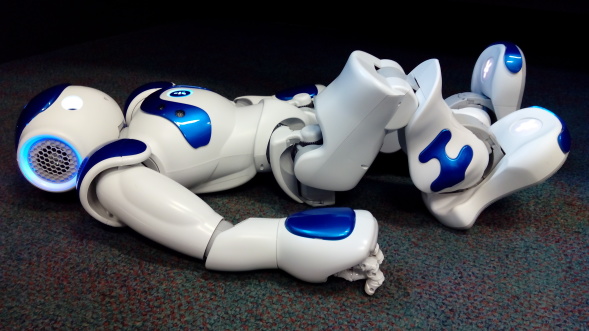} 
    \end{subfigure}
    \begin{subfigure}[b]{0.3\textwidth}
        \subcaption{\textbf{Hip Knee Flexion Lifting}}
        \includegraphics[width=\columnwidth]{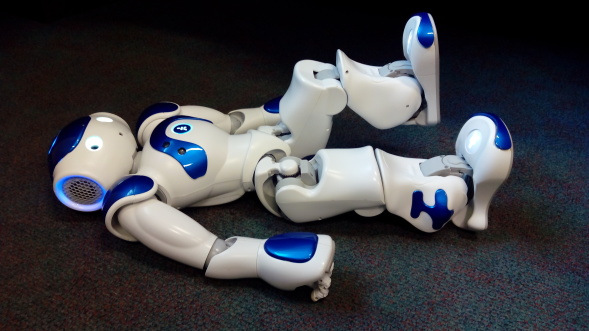} 
    \end{subfigure}
    \begin{subfigure}[b]{0.3\textwidth}
        \subcaption{\textbf{Knee Extension on Side}}
        \includegraphics[width=\columnwidth]{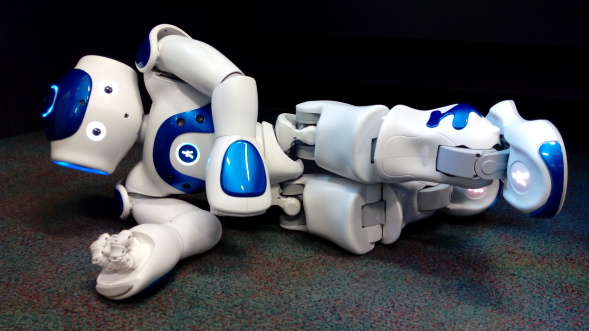}  
    \end{subfigure}

    \begin{subfigure}[b]{0.3\textwidth}
        \subcaption{\textbf{Leg Raises}}
        \includegraphics[width=\columnwidth]{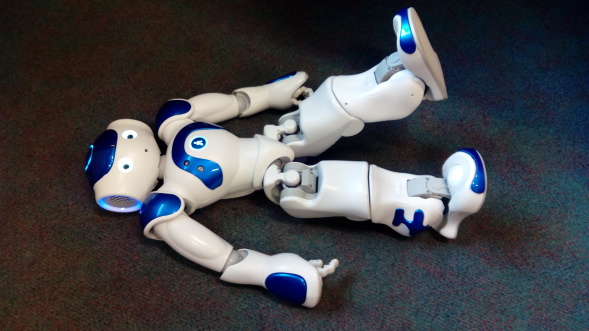} 
    \end{subfigure}
    \begin{subfigure}[b]{0.3\textwidth}
        \subcaption{\textbf{Quads over Roll}}
        \includegraphics[width=\columnwidth]{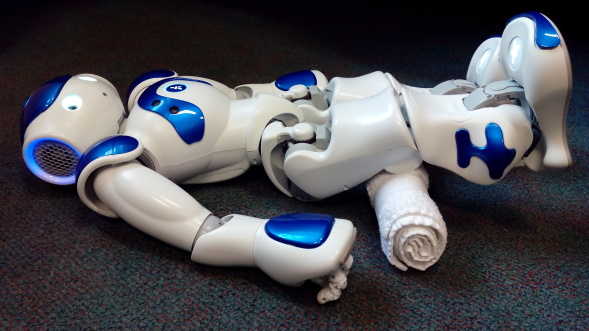} 
    \end{subfigure}
    \begin{subfigure}[b]{0.3\textwidth}
        \subcaption{\textbf{Static Quads}}
        \includegraphics[width=\columnwidth]{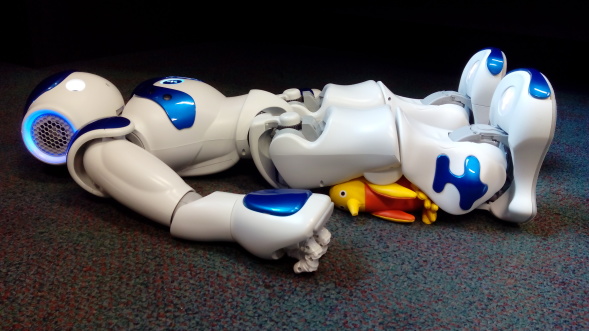}  
    \end{subfigure}

    \caption{Rehabilitation exercises executable from a lying down position. 
    \textbf{(a) Bridge:} Strengthening exercise for the hip extension muscles;
    \textbf{(b) Hip Abduction Laying:} Strengthening exercise for hip abduction muscles;
    \textbf{(c) Hip Abduction on Side:} Progression of hip abduction laying;     
    \textbf{(d) Single Bridge:} Progression of double leg bridge;    
    \textbf{(e) Hip Extension Easy:} Strengthening exercise for the hip extension muscles. This is easier than bridges and can be done with children who are not allowed to take weight through the legs;
    \textbf{(f) Hip Extension Hard:} Progression of Hip Extension Easy. Keeping the knee straight while extending the hip makes this exercise harder;
    \textbf{(g) Hip Knee Flexion Sliding:} Strengthening exercise for the hip flexors and can also be used to encourage increased range of movement at the hip and knee. The weight of the leg is supported by the bed;     
    \textbf{(h) Hip Knee Flexion Lifting:} Strengthening exercise for the hip flexors and improving range of movement at the hip and knee;
    \textbf{(i) Knee Extension on Side:} In this exercise gravity is eliminated, meaning it is an easier exercise for strengthening the muscles that extend the knee;     
    \textbf{(j) Leg Raises:} Strengthening exercise hip flexors and quadriceps;
    \textbf{(k) Quads over Roll:} Strengthening exercise for the hip extensor muscles;
    \textbf{(l) Static Quads:} This exercise is used to start practising engaging the muscles that extend the knee. It is easier than quads over roll.
    }
    \label{fig:exercises}
\end{figure}

The current Phase 2 prototype supports 13 different rehabilitation exercises:
a sit-to-stand exercise (Figure~\ref{fig:SitStands}) and 12 executable from a lying down position (Figure~\ref{fig:exercises}).
These exercises represent core lower-body exercises typically prescribed in
the rehabilitation program of patients with cerebral palsy.
Exercises have been programmed with the help of physiotherapists, through
manual positioning of the unstiffened robot to capture key postures and the temporal sequence of transitions for each 
exercise~\shortcite{mccarthy2015robots}.  This is supported using the vendor-supplied development environment, 
Choreographe~\cite{pot2009choregraphe}.

\begin{figure}[htb]
\centering
\includegraphics[trim=0 0 0 75,clip,width=0.7\columnwidth]{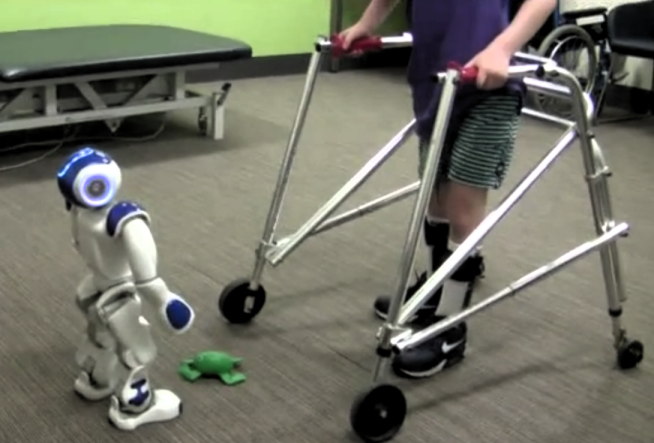}  
\caption{NAO leads a patient with cerebral palsy through the \textbf{Toy Relay} game during a therapy 
session.
[Guardian consent provided]}
\label{fig:ToyRelay}
\end{figure}

Figure~\ref{fig:ToyRelay} depicts an activity scenario in which the robot guides patients through a so-called
\emph{toy-relay} game. In this scenario, the robot asks the patient to fetch named toys on the other side of the room. 
The activity encourages patients to walk while the robot provides instructions and motivational statements.

A final supported activity scenario provides a farewell, rewarding the patient's efforts at the end of the session with a dance. 
Dance options include one programmed entirely by a physiotherapist on the research team.

\section{Design Decisions}
\label{sec:designdecisions}
The current Phase 2 prototype provides a baseline system enabling NAO to serve as an SAR for rehabilitation.  
Design requirements outlined in Section 3 have been carefully considered in the context 
of ensuring a reliable system for ongoing iterative development.  
In this section we discuss specific design choices,   compromises and considerations that have been made to meet this objective. 

\subsection{Activity Configuration Interface}
\label{subsec:ActConf}
Phase 1 required program code to be explicitly written for each session to meet the  needs of each individual patient.
However, to fulfil both \emph{Configurability} and \emph{Stand-alone} requirements, 
all  activity scenarios in the Phase 2 prototype (outlined in Section~\ref{sec:ActiScen}) 
are selectable and configurable via a text-based interface, avoiding any code modifications between sessions.
This implementation allows a session to be configured by selecting and sequencing exercises in the system, 
together with the number of sets, repetitions and execution speed.
Other parameters entered to personalise the session are the patient and the carer's name.
Configuration of the SAR is currently done via a text file edited by  a Technology Developer on behalf of the therapist.   
Development of a carer's interface is currently underway, and will soon be deployed as part of the system.

\subsection{Rehabilitation Exercises}
\label{subsec:RehabEx}
All  rehabilitation exercises and activities described in Section~\ref{sec:ActiScen} are standard exercises in existing rehabilitation programs
(\emph{Integration} requirement).
However, changes to the initial design of some exercises were required to accommodate \emph{Stability}, \emph{Robustness} and \emph{Endurance} requirements.  
For example,  the Sit-to-Stand exercise was originally designed to work with a seat, requiring pre-positioning before exercise execution. 
However, due to
an observed high risk of failure in Phase 1 (eg.,  movement of the seat or incorrect positioning), the activity was redesigned in consultation with 
therapists to incorporate  a crouching action instead.  This was more reliable and simpler to initiate.  

Walking exercise demonstrations were trialled in Phase 1, but not included in the Phase 2 prototype.
In line with Malik et al. \cite{malik2015human}, therapists deemed the crouching gait of the NAO robot 
as not appropriate for demonstration to patients.  
Furthermore,  Phase 1 highlighted  issues with both the speed and stability of NAO's walk.  
For example, the toy-relay activity scenario was designed to motivate walking in the patient by having the robot issue instructions, and through face tracking
and motivational utterances, provide
patients a sense of being monitored and encouraged during the activity
(Figure~\ref{fig:ToyRelay}).

\subsection{Activity Execution Order}
\label{subsec:ActOrder}
 It was observed during rehabilitation sessions in Phase 1 that therapists often wanted to modify the schedule of
     exercises, to better adapt to the patient's mood and energy levels.  This was easily facilitated in Phase 1 with Technology Developers in place,
but required careful consideration for Phase 2's stand-alone system.
Providing therapist's the ability to schedule the execution order of rehabilitation activities was thus deemed central to the \emph{Flexibility} requirement but
needed careful balancing with  \emph{Stability} and \emph{Endurance} requirements of the system.
For example,  while some therapists expressed a desire for on-line 
reordering of activities during sessions,  this was not incorporated into  our initial Phase 2 prototype due to increased risk of failure during
transitions between some exercise poses.   This decision was supported by  observations of care delivery in Phase 1, which 
revealed a general tendency for therapists to maintain the basic order of exercises, 
and in particular, to group exercises based on the required posture or stage of the 
session (e.g., lying down versus standing-up,  muscle strengthening versus relaxing).  
 
\subsection{Exercise Speed}
The speed of exercise execution was noted as something 
that needed to be changeable during sessions.  
Phase 1 made clear that not all patients perform exercises at the same speed, 
and during intensive rehabilitation, are likely to progress to more capable levels.  
Physiotherapists request children perform exercises at different speeds based on their clinical
observations of exercise performance. 
This may include performing some exercises faster, or slower, or holding a position for longer.   
Therefore, all the exercises have been programmed for three different speeds, 
allowing therapists the ability to select a speed during pre-configuration, 
and during the execution of an exercise set to support the \emph{Adaptability} requirement
(more details explained in subsection~\ref{subsec:TacInt}).

Static Quads is the fastest exercise in which each repetition in the fast speed setting takes 2 seconds, dropping to 
5 seconds in the slow speed setting. Hip Abduction is the slowest exercise, in which each
repetition takes 7 seconds on the fast speed setting, increasing to 15 seconds when set to slow speed.
Exercise speeds were validated based on initial observations
of the robot performing the exercises and then clinical observation of a child performing
exercises with the robot. Physiotherapists provided feedback to Technology Developers to make
speed adjustments based on this.

\subsection{Human-Robot Interaction}
\subsubsection{Robot Gestures and Speech}
\label{subsec:RobSpeech}
 Observations during Phase 1 and early testing of the Phase 2 prototype highlighted a need for speech at frequent and intermittent points to avoid long periods of silence.  In Phase 1
    this was easily accounted for through Wizard-of-Oz operations, but the \emph{Stand-Alone} requirement forced the Phase 2 
prototype to be equipped with an extensive scripted list of utterances, selected randomly, for specific activity scenarios.
Therapists suggested the inclusion  of motivational statements, as well as  reminders of important aspects of the movement to maximise therapeutic benefit.  
Motivational statements such as  \emph{``Keep it Going!''},  or \emph{``Every exercise we do gets us closer to my awesome dance moves!''} are randomly selected,
and interleaved with exercise-specific reminders  such as \emph{``Can you lift your bottom any higher?''}, or, 
\emph{``Keep your toes pointing up!''}.
Constant feedback is also provided during exercise execution by counting each repetition aloud.

Due to robustness and reliability considerations in the Phase 2 prototype, no patient progress monitoring has been incorporated into the SARs feedback
to patients. Thus,  statements are designed to be relevant to the specific exercise, but not  specific to the particular patient's current actions or progress.  
While therapist feedback made clear a desire for patient-monitoring to inform the
 delivery of  statements, this was not regarded as a prerequisite to clinical deployment.
 
Along with speech, animated gestures and actions have been incorporated into the SAR.  
Chidambaram et al. \cite{chidambaram2012designing} studied how  appropriately designed vocal and non-verbal cues can increase
compliance in people when instructed by a robot.
Accordingly, we have incorporated built-in gestures for animated speech  to enhance compliance and the overall authenticity of
interactions with patients.

\subsubsection{Speech Recognition}
\label{subsec:SpeechRec}
The challenges of speech recognition with social robots such as NAO, 
and for voice recognition  with children more generally, are well documented in the Human-Robot
Interaction literature~\cite{Kennedy2017Child}.
Pelikan and Broth~\cite{pelikan2016nao}, for example, note issues associated with the required turn-taking between robot and human when delivering speech, 
which users often find difficult to adapt to.
Challenges due to insufficient loudness of voiced responses, or unexpected statements provided by human users, 
all pose significant challenges for SARs seeking to foster natural and authentic   interactions with users.  

Phase 1 confirmed all of these issues as significant challenges, but also highlighted issues more specific to the
clinical context.  For example, errors in speech recognition would cause NAO to provide inappropriate 
responses due to misclassification of responses to questions such as  \emph{``How are you going?''}.   
Negative patient responses  were  sometimes classified as positive (and vice versa), potentially impeding the SAR's primary role as a motivator and companion.
This was exacerbated by the relatively young age of children,
and in some cases, speech impediments relating to their disability.
A lack of response to a patient's answer  would also result in long periods of silence, often requiring 
a supervising  adult to intervene and repeat the command.

Such challenges, however, were countered by Phase 1 observations that  children reacted positively 
when the robot did respond appropriately.  The incorporation of limited speech recognition  was thus  deemed important to realise  \emph{Interaction}
and \emph{Responsiveness} requirements.  To preserve  \emph{Stand-alone} and \emph{Integration} requirements of the system,
bi-directional communication was governed by specific  structural choices to constrain possible responses, and to ensure \emph{robustness} to misclassified
utterances.  These choices included:

\begin{itemize}
\item Prompting users only for simple, specific one-word verbal responses such as: \emph{``When you're ready to start, just say `go!' ''}, and/or
asking scripted questions with a constrained set of possible one-word responses (eg., Yes/No).  
\item Providing non-verbal tactile-based interaction alternatives.  For example: \emph{``Sorry, I didn't hear you! You can also tap my head to continue''}.
\item Providing speech recognition with an array of possible responses from which to base speech classification.  For example: \emph{``Yes'', ``Yeah'', ``Sure'', ``Okay'', ``Yep''}
\item Capping the waiting period for a patient response at two seconds to ensure no undue pressure was placed on the patient to provide a response. A lack of response would simply be followed by a generally relevant statement before continuing execution of the scenario.  A two second listening time was chosen from empirical observations in Phase 1.
\end{itemize}

A limited number of more open interactions were also included to allow patients the opportunity to engage more freely and express feeling and emotion (eg., \emph{``How are you going ?''}).  
Such interactions were included, in part, to allow supervising care-givers (and researchers) a chance to gauge the patient's  emotional state during the session.
SAR responses  to patient answers were designed to be generally relevant rather than response-specific.  
For example, a patient's response, either negative or positive, might be followed by the generic statement: \emph{``I am having a great time doing these exercises together with you''}.

\subsection{Visual Cues}
\begin{figure}[htb]
    \begin{subfigure}[b]{0.29\textwidth}
        \centering
        \subcaption{}
        \includegraphics[width=16mm]{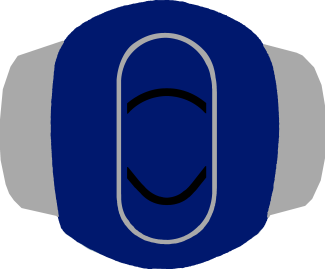} 
        \includegraphics[width=16mm]{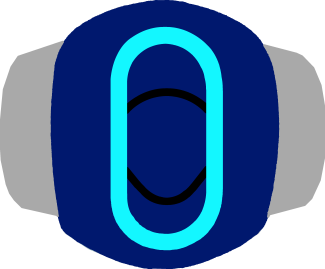} 
        \label{subfig:NormalBlink} 
    \end{subfigure}
    \begin{subfigure}[b]{0.39\textwidth}
        \centering
        \subcaption{}
        \includegraphics[width=16mm]{robotHead_Off} 
        \includegraphics[width=16mm]{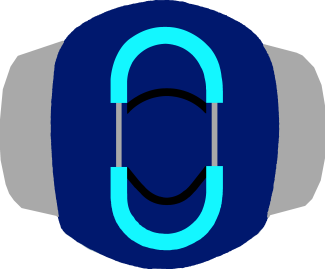} 
        \includegraphics[width=16mm]{robotHead_On} 
        \label{subfig:PausedBlink} 
    \end{subfigure}
    \begin{subfigure}[b]{0.29\textwidth}
        \centering
        \subcaption{}
        \includegraphics[width=16mm]{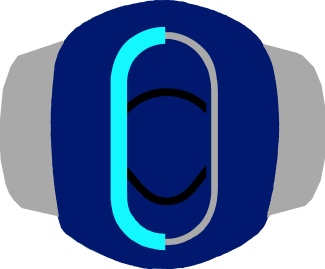} 
        \includegraphics[width=16mm]{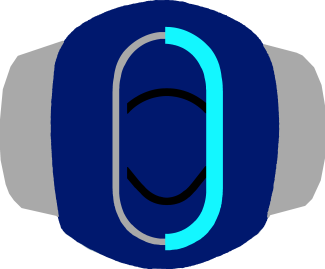} 
        \label{subfig:SideBlink} 
    \end{subfigure}

    \caption{LEDs effects (grey and thin line when LEDs are off, cyan or thick line when LEDs are on). 
    \textbf{(a) Prompting a patient/carer head-tap}
    \textbf{(b) Indicating system is paused}
    \textbf{(c) Indicating a system setup in progress}
    }
    \label{fig:LEDsEffects}
\end{figure}

To  support \emph{Interaction} and \emph{Stand-alone} requirements, NAO provides multiple LED outputs to prompt user input and convey that the system state.
LEDs around the three head-buttons of the NAO are used extensively to cue required button presses to 
 confirm progression to the next activity.  
LEDs blink at \SI{2}{Hz}, cueing the need for the head to be tapped either between exercise 
sets, or when changing activity scenarios.  
Phase 1 indicated visual cueing greatly improved the ability and confidence of people to perform the task.  
Full blinking of head LEDs is used to cue confirmation of progression to the next activity (Figure~\ref{subfig:NormalBlink}).
Other patterns of LED flashing convey the system is setting up (Figure~\ref{subfig:SideBlink}), 
or in a paused state (Figure~\ref{subfig:PausedBlink}).  

Additional LED cueing on either side of NAO's head conveys the expectation of a verbal input - most commonly 
as an alternative to head tapping for confirming progression to the next activity.

\subsection{Tactile Interface}
\label{subsec:TacInt}
\begin{figure}[htb]

    \centering  
    \begin{subfigure}[t]{\textwidth}
        \centering  
        \subcaption{Continue/Go!}
        \includegraphics[width=20mm]{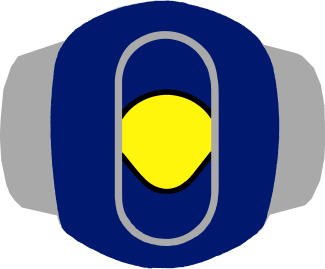} 
        \includegraphics[width=20mm]{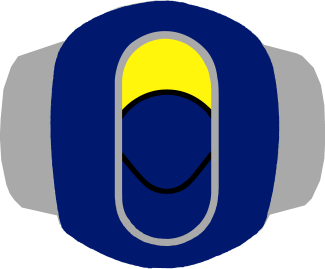} 
        \includegraphics[width=20mm]{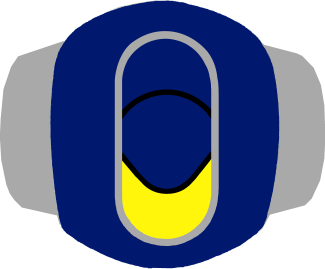} 
        \label{subfig:Singletap} 
    \end{subfigure}

    \begin{subfigure}[t]{0.5\textwidth}
        \centering  
        \subcaption{Changing Speed}
        \includegraphics[width=20mm]{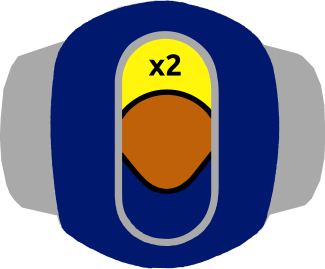}  
        \includegraphics[width=20mm]{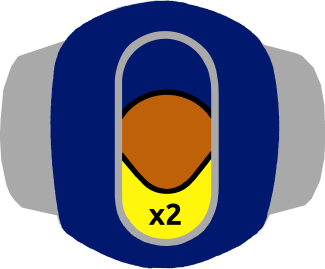} 
        \label{subfig:ChangeSpeed} 
    \end{subfigure}
    \begin{subfigure}[t]{0.4\textwidth}
        \centering  
        \subcaption{Pausing}
        \includegraphics[width=20mm]{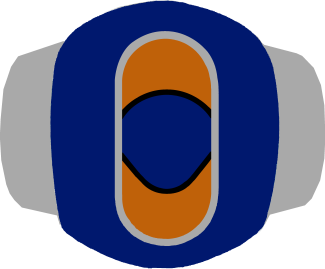} 
        \label{subfig:PauseRobot} 
    \end{subfigure}

    \caption{Tactile interface.
    \textbf{(a) Continue/Go!} a single tap on any of the three tactile buttons (middle, front, or rear) when requested by the SAR to continue;
    \textbf{(b) Changing Speed:} One finger middle button long press while second finger double tapping the front bottom to go faster, or double tapping the rear button to go slower.
    \textbf{(c) Pausing the robot:} Long press to front and rear buttons at the same time.
    }
    \label{fig:TactileInterface}
\end{figure}

Use of the NAO's head-based tactile  sensors  provides carers and patients  an alternative to speech for SAR interaction.  
In therapy sessions, patients can use the tactile interface when prompted to continue to the next activity, or to start another set of repetitions.
To ensure simplicity for patients, this is achieved  via a single tap of any of the three buttons (Figure~\ref{subfig:Singletap}).

To support online \emph{Adaptability} and \emph{Configurability} requirements, head taps were also used to provide 
carers the ability to adjust activity settings.  Most prevalent  in Phase 1 observations were scenarios in which patient performance 
required adjustment of exercise speed, or pausing of the session to accommodate unpredictable actions.

Speed adjustments are achieved using a sustained press of the NAO's middle head  touch sensor, followed by  a double tap of the front
sensor to slow down the exercise, or to the rear sensor  to speed it up (Figure~\ref{subfig:ChangeSpeed}).
To pause the robot, the rear and the front button are long pressed at the same time (Figure~\ref{subfig:PauseRobot}).
Robot adjustments are less simple than head taps to prevent re-adjustments by mistake (\emph{Robustness} requirement) 

\subsection{Human-assisted activities}
\label{subsec:Help}
 While NAO  offers a high degree of autonomy,  Phase 1 observations
highlighted limitations in the context of its ongoing therapeutic use. 
Physical constraints as well as other system uncertainties  limit the ability of the robot to perform certain exercises, 
attain certain postures, or position itself with respect to supportive auxiliary aids.
Even where autonomy may be possible, motor wear-and-tear, uncertainty of success and time costs associated 
with completing some actions autonomously motivated the use of human assistance in certain instances to meet 
 \emph{Robustness}, and \emph{Reliability}  requirements.

\begin{figure}[thb]
    \centering  
    \begin{subfigure}[t]{0.4\textwidth}
        \subcaption{Positioning the robot}
        \includegraphics[trim=30 10 70 0,clip,width=\textwidth]{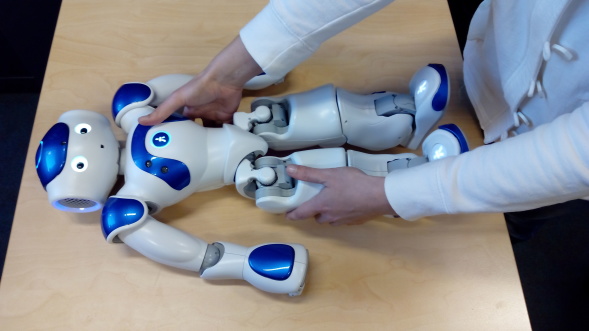}
        \label{subfig:Help1} 
    \end{subfigure}
    \begin{subfigure}[t]{0.4\textwidth}
        \subcaption{Placing auxiliary aids}
        \includegraphics[trim=50 12 75 12,clip,width=\textwidth]{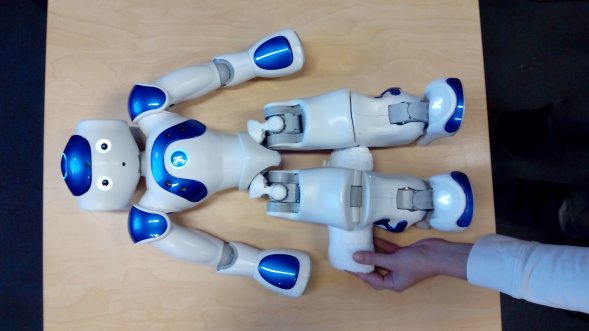}
        \label{subfig:Help2} 
    \end{subfigure}

    \begin{subfigure}[t]{0.4\textwidth}
        \subcaption{Posture}
        \includegraphics[trim=50 0 75 25,clip,width=\textwidth]{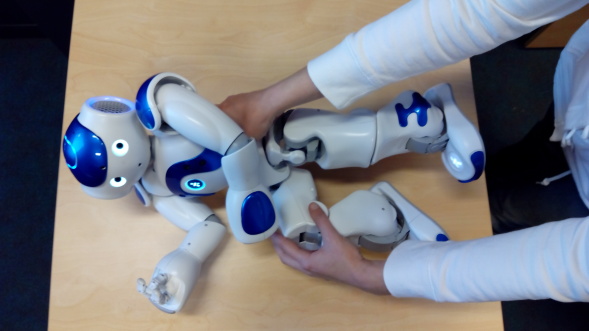}
        \label{subfig:Help3} 
    \end{subfigure}
    \begin{subfigure}[t]{0.4\textwidth}
        \subcaption{Helping to keep pace}
        \includegraphics[trim=0 0 90 0,clip,width=\textwidth]{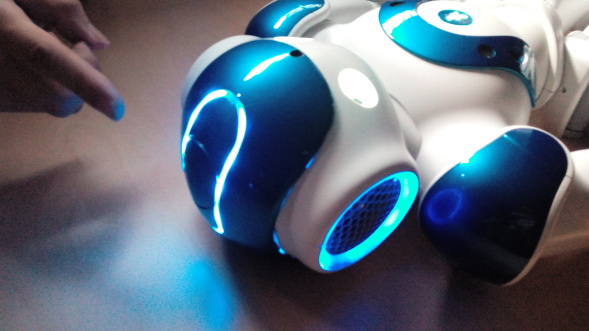}
        \label{subfig:Help4} 
    \end{subfigure}

    \label{fig:figureHelpingRobot}
    \caption{Examples of Human-assisted activities}
\end{figure}

The inclusion of robot capabilities needing human assistance, 
while unavoidable, required careful consideration. 
To meet  \emph{Integration} and \emph{Stand-alone} operation requirements, the inclusion of activity scenarios requiring carer assistance needed
to be complimentary to existing carer tasks - in particular, preserving the carer's focus on the needs of the patients.
In consultation with therapists, the following human-assisted capabilities have been implemented in the Phase 2 prototype:

\begin{description}
\item[Positioning:] 
Activity scenarios can be done in a range of different places and different positions: 
On the floor, on a table, laying down, standing up, etc. While NAO can stand-up or lay down by itself,
manual re-positioning, in which the therapist lifts and places the robot close 
to the patient, is quicker, less error-prone, and reduces wear-and-tear (Figure~\ref{subfig:Help1}) than having the robot position itself.

\item[Placing auxiliary aid:]  
\emph{Quads over Roll} and \emph{Static Quads} are the two exercises where, as with the patient, a small rolled 
towel is placed under the leg of the robot (Figure~\ref{subfig:Help2}).
The robot will ask explicitly for this kind of assistance:

\emph{``For Quads over Roll we will need to roll two towels. One big for you, 
and a little one for me! We have to put the towel under our left knee.''}

\item[Posture:] 
\emph{Hip Abduction on Side}, \emph{Hip Extensions}, and \emph{Knee Extension on Side} 
are exercises where the robot needs to be rolled onto its side (Figure~\ref{subfig:Help3}).
Like with auxiliary aids, the robot asks explicitly for this kind of assistance:

\emph{``For this exercise, I will need your help!   I will need you to roll me onto my right side. Can you do that for me?''}

\item[Keeping pace:] 
Between exercises the SAR lets the patient rest. A head-tap (Figure~\ref{subfig:Help4})
is used to indicate progression to the next activity.
Head-taps are also used to confirm progress during instructional activities such as \emph{Sit-to-Stands} or \emph{Toy Relay}.

\emph{``Say Go! Or tap my head when you are ready to start the next set''}

\end{description}

Our preliminary results showed that the amount of time physiotherapists had to deal with the robot did not negatively 
impact patient sessions \cite{marti2016help}. 
Phase 2 evaluation is closely examining time-costs and frequency of such requests with respect 
to the overall perceived benefits of the system.  We discuss this further in Section~\ref{sec:DesignEvaluation}.

%
\section{Initial Clinical Deployment :  System and In-situ Design Critique}
\label{sec:clinicalsessions}
Initial Phase 2 testing has commenced in preparation for a planned clinical evaluation of the system.   In this section we present
our methodology for evaluating the Phase 2 system, as we develop the SAR for clinical trials. Here 
we focus primarily on operational aspects of the SAR with respect to the 
requirements and design decisions outlined in previous sections.  We also present initial user response data from therapists and parents
who have observed 
the SAR in the clinical care of patients.   
Due to the early stage of clinical testing, 
we defer a comprehensive evaluation of the SAR's perceived 
therapeutic benefits and patient/parent/therapist perceptions until the completion of Phase 2 testing.
 
\subsection{Phase 2 Testing Setup}
\begin{figure}[th]
\centering
\includegraphics[trim=110 350 122 150,clip,width=0.7\columnwidth]{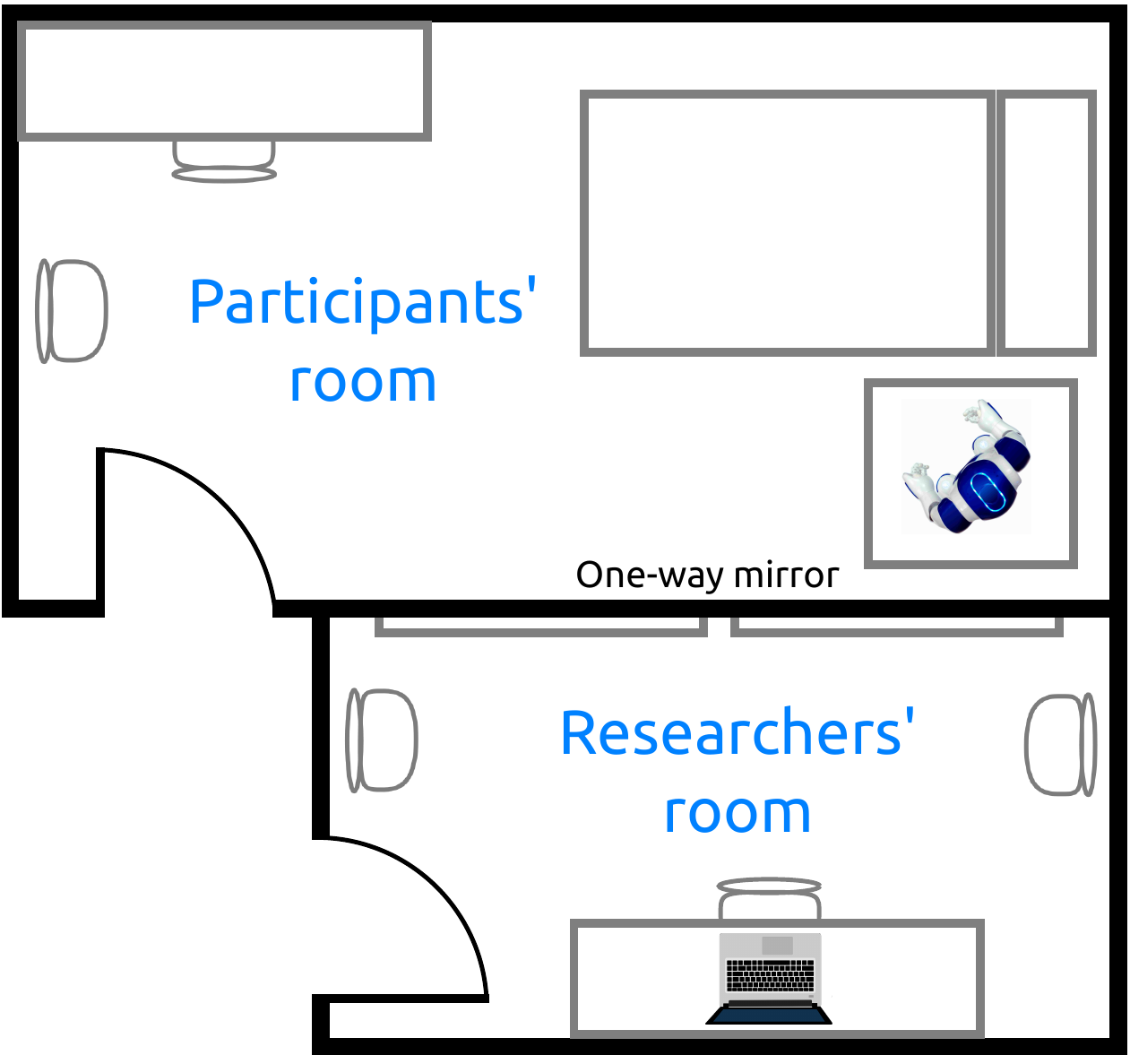}  
\caption{Study setting floor plan.}
\label{fig:FloorPlan}
\end{figure}

Phase 2 clinical sessions with the robot are conducted in a consultation room at the
rehabilitation clinic of the Royal Children's Hospital, Melbourne, Australia. 
Observing investigators reside in an adjacent observation room  with one-way mirror
(see Figure~\ref{fig:FloorPlan}).  
The patient, a therapist and the SAR are in the Participants' room. 
Parents can observe the session from either of the two rooms.
All participants are informed that sessions are being observed by research team members.  
Pre-configuration of the system is performed by a research team member.
Configuration options are communicated to the research team member by the 
treating therapist prior to each session. 
 
Before starting the session, the robot is placed in a crouched position on a table-top 
next to the bed and the attending therapist receives a 5 minute informal introduction to the system. 
In this introduction it is explained that the robot will work autonomously, 
will be able to recover from some failures, however may ask for help for particular positioning requirements,
or request head-taps to confirm session progression.

The session starts with the robot greeting the patient and introducing itself.
NAO then commences the patient's 
pre-configured exercise program as described in Section~\ref{sec:ActiScen}.

NAO's software currently runs off a laptop with wireless connection to the robot.
During each session, an attending research engineer monitors the software in
the adjacent observation room, and interacts with the system only if necessary (ie., a system failure requiring a reset of the system).
All operational requirements 
are thus handled by attending care-givers and the patient.  
Our protocol allows engineer intervention to occur only when a
system error or issue is disrupting the session, and is easily recoverable in-situ.  All such instances are logged.  

\subsection{Data collection}
\label{subsec:Data}
During each session, observations on 
the system performance, usability, and interactions 
among participants and the robot are recorded by observing research team members.
System logs for each session are also recorded, capturing
exercise configuration, completed exercises (by the robot), user prompts, 
number of requests for help, and time required for needs to be met.

A key focus of Phase 2 testing is the evaluation of the SAR's perceived utility, ease-of-use, and participants' trust of
the robot as a therapeutic device.  
To this end, survey response data is collected using adapted versions of the robot acceptance questionnaire originally proposed by
Heenrik et al.~\cite{heerink2009measuring}.  Responses are recorded using a 
Likert scale, with specific versions of the questionnaire used for each of 
the three participant groups (patients, parents, therapists).   

Tables~\ref{tab:QPT_results} and \ref{tab:QG_results} present the adapted survey questions for   
physiotherapists and parents respectively, along with initial responses (discussed further below).
The questionnaire is divided into different categories: Anxiety (ANX1, ANX2), 
Attitude (ATT), Facilitating Conditions (FC), Intention to Use (ITU), 
Perceived Adaptability (PAD), Perceived Ease of Use (PEOU), Perceived Usefulness (PU), Trust (TR), 
and Social Influence (SI). 
Anxiety category is divided in two parts to better understand the extent to which participants were anxious
about their safety with the robot as distinguished from anxieties associated with using the system correctly and without damaging it.

Participants are also asked open questions seeking feedback on strengths and weaknesses of the system, desirable features currently
not present, and their impressions of trust and benefit.  Due to the young age and cognitive impediments of many of the patients
expected to be recruited in Phase 2 testing,  only limited survey feedback is expected from patients, as determined by 
physiotherapist clinical judgement.


\subsection{Preliminary Session Results} \label{subsec:SystemEvaluation} 
\subsubsection{System Performance}
At the time of writing, our Phase 2 prototype has led 14 observed sessions of up to 30 minutes each.
Table~\ref{tab:Sessions} 
provides a structured overview of the 14 sessions, indicating which exercises were performed, the duration of each session, exercises completed, and any system disruptions that may have occurred.

\begin{table}[htbp] \caption{Rehabilitation sessions in Phase 2 summarised. Fourteen sessions, with nine different patients. 
                            The exercises programmed are chosen by the patient's physiotherapist. Duration of the rehabilitation session
                            including introductory speech and farewell dance in mm:ss format.}~\label{tab:Sessions} \centering 
\scalebox{0.7}{
    \begin{tabular}{ c c l l c p{4cm} p{4cm} } \toprule
    No  & Patient & Exercises programmed   & Exercises completed  & Duration &  System Disruptions & Comments \\ 
    \midrule 
    1   & P-1 & Quads over Roll            & Quads over Roll        & N/A   &                           & Patient expressed positive    \\   
        &     & Bridge                     & Bridge                 &       &                           & attitudes and showed focus    \\   
        &     & Hip Abductions Laying      & Hip Abductions Laying  &       &                           & on the SAR.                   \\
    \midrule 
    2   & P-2 & Quads over Roll            & Quads over Roll        & N/A   & The last exercise was not & The patient did not like      \\   
        &     & Bridge                     & Bridge                 &       & executed due to an error  & the robot.                    \\
        &     & Hip Knee Extension         &                        &       & in the system.            &                               \\
    \midrule 
    3   & P-3 & Hip Knee Flexion Lifting   &                        & N/A   &                           & Session aborted. Robot was    \\   
        &     & Toy Relay                  &                        &       &                           & too loud, upset the patient.  \\
    \midrule 
    4   & P-4 & Static Quads               & Static Quads           & 19:41 &                           & Patient proactively helped    \\   
        &     & Quads over Roll            & Quads over Roll        &       &                           & the robot when required.      \\   
        &     & Single Bridge              & Single Bridge          &       &                           & Due to patient's fatigue,     \\   
        &     & Hip Knee Flexion Sliding   & Hip Knee Flexion Sliding &     &                           & physiotherapist shortened     \\   
        &     & Toy Relay                  &                        &       &                           & the session. Patient happy to \\
        &     &                            &                        &       &                           & do another session.           \\
    \midrule 
    5   & P-5 & Quads over Roll            & Quads over Roll        & 10:50 &                           & Patient showed enthusiasm     \\   
        &     & Bridge                     & Bridge                 &       &                           & for a session with the SAR.   \\   
        &     & Hip Abduction on Side      &                        &       &                           & Patient's frustration         \\
        &     & Leg Raises                 &                        &       &                           & with the exercises shortened  \\
        &     &                            &                        &       &                           & the session.                  \\
    \midrule 
    6   & P-4 & Toy Relay                  & Toy Relay              & 23:23 &                           &                               \\
    \midrule 
    7   & P-6 & Static Quads               & Static Quads           & 16:19 &                           & Patient expressed positive    \\   
        &     & Hip Abductions Laying      & Hip Abductions Laying  &       &                           & attitudes towards the robot,  \\
        &     & Toy Relay                  & Toy Relay              &       &                           & enjoyment and excitement.     \\   
    \midrule 
    8   & P-6 & Static Quads               & Static Quads           & 24:52 &                           & Patient showed focus on the   \\   
        &     & Quads over Roll            & Quads over Roll        &       &                           & the robot. Patient happy to   \\   
        &     & Leg Raises                 & Leg Raises             &       &                           & do another session.           \\
        &     & Toy Relay                  & Toy Relay              &       &                           &                               \\
    \midrule 
    9   & P-6 & Static Quads               & Static Quads           & 25:42 &                           & Patient expressed positive    \\   
        &     & Quads over Roll            & Quads over Roll        &       &                           & attitudes towards the robot,  \\
        &     & Leg Raises                 & Leg Raises             &       &                           & smiled and interacted with    \\   
        &     & Toy Relay                  & Toy Relay              &       &                           & robot.                        \\
    \midrule 
   10   & P-7 & Sit-to-Stands              &                        & N/A   &                           & Session aborted. Patient      \\
        &     & Toy Relay                  &                        &       &                           & non-compliant in therapy      \\
        &     &                            &                        &       &                           & sessions.                     \\
    \midrule
   11   & P-8 & Static Quads               & Static Quads           & 17:15 &                           & Patient showed enjoyment      \\
        &     & Quads over Roll            & Quads over Roll        &       &                           & and proactively helped the    \\
        &     & Leg Raises                 & Leg Raises             &       &                           & robot when required.          \\
    \midrule
   12   & P-8 & Static Quads               & Static Quads           & 16:55 &Robot fall during the final& Patient showed focus on the   \\
        &     & Quads over Roll            & Quads over Roll        &       &dance routine. No technical& robot. Patient happy to       \\
        &     & Leg Raises                 & Leg Raises             &       &intervention was required. & do another session.           \\
    \midrule
   13   & P-8 & Static Quads               & Static Quads           & 16:50 &                           & Patient showed focus on the   \\
        &     & Quads over Roll            & Quads over Roll        &       &                           & robot and expressed positive  \\
        &     & Leg Raises                 & Leg Raises             &       &                           & attitudes when interacting.   \\
    \midrule
   14   & P-9 & Static Quads               & Static Quads           & 31:35 & Battery drainage.         & Teenager patient liked the    \\
        &     & Quads over Roll            & Quads over Roll        &       & Engineer intervention was & experience, but preferred to  \\
        &     & Bridge                     & Bridge                 &       & required to restart the   & do rehabilitation with a      \\
        &     & Hip Abductions Laying      & Hip Abductions Laying  &       & system.                   & physiotherapist to have a     \\
        &     & Hip Knee Flexion Sliding   & Hip Knee Flexion Sliding &     &                           & proper conversation.          \\
        &     & Sit-to-Stands              & Sit-to-Stands          &       &                           &                               \\
    \bottomrule

    \end{tabular}
}
\end{table}

Of the 14 sessions conducted, 9 sessions finished with patients completing all prescribed exercises.  
Two of the five sessions not completed fully were shortened by the attending physiotherapist 
(Sessions 4 and 5) based on clinical judgement. In Session 4, the Toy Relay was excluded due to patient fatigue (though the patient 
remained positive throughout the session), and in Session 5, two programmed exercises were not 
conducted due to the patient's perceived lack of stamina.  One session was aborted due to an unrecoverable system error (Session 2), 
causing the last prescribed exercise for the session to be completed without the robot.  

Sessions 12 and 14 involved recoverable system disruptions.  In Session 12, a loss of stability occurred during the final 
dance behaviour, and in Session 14, a loss of power (after back-to-back sessions) required engineer-intervention to resolve.

Two sessions were aborted due to patients' explicit expression of dislike of the SAR.  In Session 3, a young 3 year old patient expressed 
fear of the robot due to its loudness, causing an immediate halt to the session. The second case, Session 10, a teenage patient
expressed a clear dislike of the robot, invoking a premature stop to the session.  Therapist feedback noted the patient has a
history of non-compliance in therapy sessions.
These events reflect a clear diversity of patient needs,
 and are informative to future development and testing of the system.

\subsubsection{Therapist/Parent Feedback}

In early Phase 2 testing, survey responses from 4 different physiotherapists have been recorded upon completion of their first 
session interacting with the SAR.  
As participant numbers are small, we present the raw quantitative data provided in Table~\ref{tab:QPT_results}, 
and an overview of the open question responses.

Inspection of these early survey responses show that 3 of the 4 recruited physiotherapists perceived the system as easy-to-use.  
However, the fourth physiotherapist (PT-3) expressed mostly neutral opinions of the system's usability, and disagreement about 
having enough knowledge of the robot to make use of it effectively.   Notably, physiotherapists had only a brief introduction to 
the SAR at the beginning of their first  session with the SAR, however, were observed to exhibit competence interacting with and 
operating the SAR.
 
All physiotherapists expressed  positive attitudes towards using the robot in rehabilitation therapy (ATT).  In response to questions of
the SAR's perceived usefulness (PU), all therapists expressed either Agreement or Strong Agreement that the robot is convenient and
useful for paediatric rehabilitation (PU).   

Responses to questions of intention to use (ITU) the SAR in future sessions presents 
a less clear picture from early data collection.  While 3 out of 4 therapists agree they would think to use the SAR during the next 
therapy session, two of these therapists respond only neutrally to being certain of this.  While no specific feedback elaborating on
these responses was obtained, it is likely that confounding factors  such as the unknown rehabilitation needs of future patients 
quite reasonably attenuated their certainty.

Physiotherapist responses to statements of trusting (TR) the robot's advice were, with the exception of one response (Agree),
either neural or in disagreement. No specific feedback was obtained 
to better understand these responses.  We discuss this further below.

When asked about the most useful features of the robot, physiotherapists reported the SAR's ability to demonstrate exercises to 
patients, and its motivational role in keeping the patient focused on each exercise as most useful. However, therapists also noted
deficits in the system's performance, including the SAR's lack of responsiveness to patient mood and performance, and battery life
in the context of back-to-back sessions (Session 14). Physiotherapists' reactions to the robot's lack of responsiveness suggests
they had expectations that the SAR would respond to the patient's mood.  However, no physiotherapist explicitly expressed the 
desire for a feature to manually change the robot's behaviour to match the patient's current state.

\begin{table}[hbt] \caption{Acceptance questionnaire for physiotherapists with their initial responses.
                            The questionnaire is dived by different constructs:
                            Anxiety (ANX1, ANX2), Attitude (ATT), Facilitating Conditions (FC), Intention to Use (ITU), Perceived Adaptability (PAD), 
                            Perceived Ease of Use (PEOU), Perceived Usefulness (PU), Trust (TR), and Social Influence (SI). 
                            Likert scale: 1 = Strongly Disagree; 2 = Disagree; 3 = Neutral; 4 = Agree; 5 = Strongly Agree.}~\label{tab:QPT_results} \centering 
\scalebox{0.75}{
    \begin{tabular}{ l l p{8cm}  c c c c } \toprule
         &    &    & \multicolumn{4}{c}{Responses} \\
    \cmidrule(r){4-7}
    Construct & No & Question & PT-1 & PT-2 & PT-3 & PT-4 \\ 
    \midrule 
    ANX1 &  1 & I would be afraid to make mistakes using the robot                                      & 1 & 2 & 3 & 3 \\ 
         &  2 & I would be afraid to break something when using the robot                               & 3 & 2 & 4 & 5 \\ 
    \midrule 
    ANX2 &  3 & I find the robot scary                                                                  & 1 & 2 & 1 & 1 \\ 
         &  4 & I find the robot intimidating                                                           & 1 & 2 & 2 & 1 \\ 
    \midrule 
    ATT  &  5 & I think it's a good idea to use the robot                                               & 5 & 4 & 4 & 4 \\ 
         &  6 & The robot would make therapy sessions more interesting                                  & 5 & 4 & 4 & 4 \\ 
    \midrule 
    FC   &  7 & I have everything I need to make good use of the robot                                  & 4 & 3 & 3 & 4 \\ 
         &  8 & I know enough of the robot to make good use of it                                       & 4 & 3 & 2 & 3 \\ 
    \midrule 
    ITU  &  9 & If I have access to the robot, I think I'll use it during the next therapy sessions     & 4 & 4 & 3 & 4 \\ 
         & 10 & If I have access to the robot, I am certain to use it in the next therapy sessions      & 4 & 3 & 3 & 3 \\ 
         & 11 & If I have access to the robot, I'm planning to use it during the next therapy sessions  & 4 & 4 & 3 & 3 \\ 
    \midrule 
    PAD  & 12 & I think the robot can be adaptive to what I need                                        & 3 & 4 & 2 & 2 \\ 
         & 13 & I think the robot will only do what I need at that particular moment                    & 3 & 3 & 2 & 4 \\ 
         & 14 & I think the robot will help me when I consider it to be necessary                       &   & 4 & 3 & 4 \\ 
    \midrule 
    PEOU & 15 & I think I will know quickly how to use the robot                                        & 5 & 4 & 3 & 5 \\ 
         & 16 & I find the robot easy to use                                                            & 5 & 4 & 3 & 4 \\ 
         & 17 & I think I will be able to use the robot without any help if I have been trained         & 5 & 4 & 3 & 4 \\ 
         & 18 & I think I will be able to use the robot when there is someone around to help me         & 5 & 4 & 4 & 5 \\ 
         & 19 & I think I will be able to use the robot  when I have a good manual                      & 5 & 4 & 3 & 5 \\ 
    \midrule 
    PU   & 20 & I think the robot is useful to help in paediatric therapy                               & 5 & 4 & 4 & 4 \\ 
         & 21 & It would be convenient to have the robot for therapy sessions with children             & 5 & 4 & 4 & 4 \\ 
         & 22 & I think the robot can help me with many things during paediatric sessions               & 4 & 4 & 3 & 4 \\ 
    \midrule 
    SI   & 23 & I think the staff would like me using the robot                                         & 3 & 3 & 4 & 4 \\ 
         & 24 & I think parents would like me using the robot                                           & 5 & 4 & 4 & 3 \\ 
         & 25 & I think patients would like me using the robot                                          & 5 & 4 & 3 & 4 \\ 
         & 26 & I think it would give a good impression if I should use the robot                       & 4 & 3 & 3 & 4 \\ 
    \midrule
    TR   & 27 & I would trust the robot if it gave me  advice                                           & 3 & 3 & 2 & 2 \\ 
         & 28 & I would follow the advice the robot gives me                                            & 3 & 4 & 3 & 2 \\ 
    \bottomrule 

    \end{tabular} 
}
\end{table}

As per our in-situ design process, feedback from parents/guardians was also sought as part of preliminary Phase 2 testing.  Raw 
survey responses for parents (N=4) attending therapy sessions are presented  in Table~\ref{tab:QG_results}.
Notably, all parents expressed overwhelming agreement to statements reflecting the SAR's Perceived Usefulness (PU).  All strongly
agreed that the SAR  is useful in their child's therapy, and all agreed the robot can help their child with many things.  
Parents also reported positive attitudes (ATT) to using the SAR in their child's rehabilitation therapy.   
Of particular interest for future testing  of the SAR is parent's perceptions of the SAR's usability (PEOU).   Three parents 
responded positively to the robot being easy to use,  and to feeling confident in using the system themselves.  One parent  (G-4)
 expressed mostly neutral responses to PEOU questions, although also strongly disagreed to being able to use the SAR without any
help.  Notably, all except one parent also expressed  disagreement or neutrality about having enough knowledge to make  good use of
the SAR.  As future testing of the SAR intends to allow parents to operate the SAR without therapist supervision, these results
are both encouraging and informative,  indicating that with more targeted training and familiarisation,  it is reasonable to expect
parents to feel capable and comfortable operating the SAR on their own.
In contrast with the physiotherapist's responses, parents in general Strongly Agree that they would Trust (TR) and follow the robot's advice.

In open feedback, 3 out of the 4 parents specifically noted the robot helped keep their child focused on completing the exercises.  
These statements included:
\emph{``[Daughter] seemed to respond really well and her mind was taken off with the robot''}; and
\emph{``The robot was useful because it had my child's attention the whole time''}.
Observational data corroborated these perceptions, with patients exhibiting high focus on the SAR during the rehabilitation session.
Notably, two parents of female patients noted they would prefer gender-neutral colouring for NAO.

\begin{table}[hbt] \caption{Acceptance questionnaire for guardians with initial responses. 
                            The questionnaire is dived by different constructs:
                            Anxiety (ANX1, ANX2), Facilitating Conditions (FC), Attitude (ATT), 
                            Perceived Ease of Use (PEOU), Perceived Usefulness (PU), Trust (TR). 
                            Likert scale: 1 = Strongly Disagree; 2 = Disagree; 3 = Neutral; 4 = Agree; 5 = Strongly Agree.}~\label{tab:QG_results} \centering 
\scalebox{0.75}{
    \begin{tabular}{ l l p{8cm}  c c c c } \toprule
         &    &    & \multicolumn{4}{c}{Responses} \\
    \cmidrule(r){4-7}
    Construct & No & Question & G-1 & G-2 & G-3 & G-4 \\ 
    \midrule 
    ANX1 &  1 & I would be afraid to make mistakes using the robot                                                              & 2 & 5 & 1 & 3 \\ 
         &  2 & I would be afraid to break something when using the robot                                                       & 1 & 3 & 5 & 4 \\ 
    \midrule 
    ANX2 &  3 & I find the robot scary                                                                                          & 1 & 5 & 1 & 1 \\ 
         &  4 & I find the robot intimidating                                                                                   & 1 & 1 & 1 & 1 \\ 
    \midrule 
    FC   &  5 & I have everything I need to make good use of the robot                                                          & 3 & 5 & 2 & 3 \\ 
         &  6 & I know enough of the robot to make good use of it                                                               & 3 & 4 & 1 & 2 \\ 
    \midrule 
    ATT  &  7 & I think it's a good idea to use the robot                                                                       & 5 & 5 & 5 & 5 \\ 
         &  8 & The robot would make my child's rehab sessions more interesting                                                 & 5 & 5 & 5 & 4 \\
    \midrule 
    PEOU &  9 & I think I will know quickly how to use the robot                                                                & 5 & 5 & 5 & 3 \\ 
         & 10 & I find the robot easy to use                                                                                    & 5 & 5 & 5 & 3 \\ 
         & 11 & I think I can use the robot without any help                                                                    & 5 & 4 & 5 & 1 \\ 
         & 12 & I think I can use the robot when there is someone around to help me                                             & 5 & 5 & 5 & 3 \\ 
         & 13 & I think I can use the robot when I have a good manual                                                           & 5 & 5 & 5 & 4 \\ 
    \midrule 
    PU   & 14 & I think the robot is useful for paediatric rehabilitation                                                       & 5 & 5 & 5 & 5 \\ 
         & 15 & It would be convenient to have the robot for therapy sessions together with the physiotherapist                 & 3 & 5 & 5 & 4 \\ 
         & 16 & It would be convenient to have the robot for therapy sessions when the physiotherapist is not in the session    & 4 & 5 & 5 & 3 \\ 
         & 17 & I think the robot can help my child with many things                                                            & 5 & 5 & 5 & 4 \\ 
    \midrule 
    TR   & 18 & I would trust the robot if it gave me advice                                                                    & 4 & 5 & 5 & 5 \\ 
         & 19 & I would follow the advice the robot gives me                                                                    & 5 & 5 & 5 & 4 \\
    \bottomrule 

    \end{tabular} 
}
\end{table}

\subsubsection{Summary}
Preliminary in-situ testing indicates the system  is performing strongly on key metrics of acceptance in clinical practise; in particular Perceived Usefulness and Perceived Ease-Of-Use.  While session observations have highlighted areas of improvement
for the system, discussed in the next section, therapists and parents respond almost universally positively to statements reflecting
the SAR's usefulness and usability.

Preliminary results on trust provide less clarity.  Indeed, the issue of trust in human-robot interaction 
research is known to be complex, and often difficult to interpret.  
Our survey responses reflect a level of distrust from therapists with respect to taking advice from the SAR.  
Certainly, the short exposure time of the therapists to the SAR is a likely factor, however it should also be noted that the 
survey questions refer only to the participant's own trust and willingness to follow the advice of the SAR, as separate from their
 trust in the system as a therapeutic aid.  That therapists express more positivity towards statements reflecting their 
Intention-to-Use the SAR in future sessions provides some support for trust of the system  in this respect.
Notably, parents express a much higher degree of trust in the SAR's advice, however it must be noted that this is likley to be
conflated with trust they may feel towards the therapist's
clinical judgement to include the SAR in their child's therapy, as well as he hospital's judgement in allowing the study to take place.
It is thus difficult to draw any clear conclusions on trust from this data.  

The robot has  successfully led 9 of the 14 sessions through the same exercises that would usually be performed as per their rehabilitation program.
Of the 5 non-completed sessions, only one was due to an unrecoverable system error.   
While such incidents are undesirable, they are easily addressed, and indeed vindicate the in-situ testing phase prior to clinical trials we are undertaking. 

Overall, preliminary observational and therapist feedback supports the assertion that the SAR is, more often than not, 
positively impacting the motivation of children to complete each exercise in full, and correctly.  
However, more data is required to thoroughly evaluate this.  
Motivation to complete independent exercises is a
known issue in physiotherapy practice (for both adults and children) and technological supports are showing promising results in
improving compliance \cite{LAMBERT2017App}.   The particular value added by an embodied artificial agent such as NAO versus a 
virtual graphical body (for example, a video or animation) has also been previously explored, 
with evidence suggesting participants perceive more social presence when interacting with a robot than with 
other virtual agents \cite{Lee2006Embodiment,mataric2007Embodiment}, which in turn may lead to heightened motivation and 
emotional connection with the aid.  Our own observations in both phases of design and development support this, with younger  
patients in particular exhibiting behaviours suggesting they believe the robot is listening and responding to them.

\subsection{System Design Evaluation}
\label{sec:DesignEvaluation}
Below we discuss and critique specific design decisions (outlined in Section~\ref{sec:designdecisions}) 
based on the early Phase 2 testing outlined above.   We discuss these in the context of developing the SAR for full clinical deployment.

\subsubsection{Configurability}
The SAR software was designed  to support rapid configuration for new exercise sessions,
allowing for the pre-selection and scheduling of exercises to perform, number of repetitions, speed of execution, 
entertainment modules, as well as patient and physiotherapist names.  
Configuration time was observed to take no more than 5 minutes, however, 
the current interface is text file based and thus not directly usable by therapists. 
While therapists were able to effectively communicate the session schedule to engineers via a text-based template,
this was an inefficient process, and will not scale to the ongoing clinical deployment of multiple robots, or multiple patients
with the same robot.
To this end, a tablet-based interface for therapists and carers is under development allowing session histories to be stored, and 
importantly, the removal of research team members from the configuration process.

\subsubsection{Stability and Robustness versus Flexibility}
The decision to fix the activity execution order during sessions was chosen to maintain  \emph{Stability} and \emph{Robustness} requirements
of the SAR by minimising posture and position changes.  
The low number of recorded system failures in Phase 2 testing supports this decision, 
with system failures to date only occurring during a dance (entertainment) scenario, 
due to a system error that has been fixed, and due to power loss (see Table~\ref{tab:Sessions}).
However,  \emph{Flexibility} is compromised, and the inability to dynamically change exercise execution order was raised as a deficit of the current system design  by therapists. 
One therapist suggested  the robot could ask the patient which exercise to do next, instead of following a prescribed order.  
Such flexibility is being considered within particular exercise subsets.   
For example, the system may allow therapists (or patients) to change execution order within a specific  block of lower body exercises.

The SAR  provides therapist's the ability to dynamically alter the robot's exercise execution speed via a simple tactile interface, however, no 
recorded instances of its use were observed over the 14 Phase 2 sessions.   Therapists have raised no specific concerns with the tactile interface, 
and  were observed to use this interface for other tasks such as  confirming progression to the next activity.  
Future work will focus more specifically on understanding the usability needs of this feature.

\subsubsection{Speech and interaction}
No specific feedback about the animated speech was provided by participants, however,  
general observations of patient reactions suggested the animated speech enhanced the SAR's authenticity with patients.
The prototype has 20 pre-programmed phrases to encourage and motivate patients common for all exercises.   
Five specific instructional phrases are also programmed for each exercise and selected randomly.
As noted previously, the system does not provide explicit monitoring and thus any detailed feedback to patients regarding their
performance is still assumed to be delivered by the physiotherapist or parent who is present.
However, the feedback from NAO was sometimes re-affirmed  by the parent or therapist to encourage the child.  For example,
one therapist said \emph{``See!  NAO is also asking you to lift your bottom higher''}. 

Both developers and therapists noted a high degree of repetitiveness in the SAR's delivered statements, 
suggesting the  range of motivational phrases should be increased.  
However, this repetitiveness was not observed to impact negatively on  engagement or compliance with the mostly young patients.    
It is interesting to  note that therapists often deliver similarly frequent repeated statements  as  a means of reinforcing positive and important feedback.   
However, such phrases are typically short and to the point.  
That the SAR is regarded as repetitive by therapists suggests it may be impeded by not just an insufficient number of  unique phrases, but also by the choice of phrases 
being repeated, or the lack of variability in their delivery.

On occasions the SAR's speech was observed to cause confusion or mild irritation in patient responses.
For example, 
NAO's counting of exercise repetitions was observed to occasionally confuse patients when not in-sync with their own perception of progress.  
Word pronunciation was also observed to be important. For example,
while most patients visibly expressed satisfaction in the SAR referring to them by name,  
incorrect pronunciations were  observed to evoke negative patient responses.  
One patient, for example, noted:
\emph{``I would like the robot to say my name correctly''}.  
Such observations in patient reaction and performance, while highlighting clear need for improvement,  do also confirm
the importance of robot speech in the SAR's design.  
Understanding how speech can be designed to best compliment the 
roles of the SAR in such therapeutic contexts is an important area of future work.
 
Despite design decisions to optimise the robustness of NAO's built in speech recognition (see Section~\ref{subsec:SpeechRec}), 
verbal interaction with the SAR  remained problematic.  
Notably, recent studies have highlighted specific issues with the NAO platform's speech recognition \cite{pelikan2016nao}, 
as well as natural language processing with children more generally \cite{Kennedy2017Child}.  
Phase 2 session observations noted frequent false negative responses to simple phrases such as \emph{`Go`'}.  
This was observed especially with patients, but also with therapists.    
The provision of alternative modes of interaction allowed sessions to continue regardless.  
Notably,  participants were observed to quickly discard verbal communication  (typically after the first failed attempt) in favour of tactile button pressing to respond.   
Providing feedback to  participants when  speech was not recognised was observed to alleviate confusion and frustration,  allowing participants 
to solve the situation themselves.

As noted, tactile button taps were observed to provide a reliable and preferred mode of interaction for both patients and therapists with the SAR.
The inclusion of  flashing LEDs marking the boundary of  the head buttons was observed to reduce  errors in precision, 
and confusion caused by missed taps observed in Phase 1.  
In particular, the continued flashing of the  LEDs until a tap was registered provided sufficient guidance to participants to make
another attempt if required, further supporting the SAR's \emph{Integration} in the session, and \emph{Stand-alone} operation.

\subsubsection{Human-Assisted Activities}
\label{sec:HumanAssistActivities}
\begin{figure}[htb]
\centering

    \begin{subfigure}[t]{0.85\textwidth}
        \centering
        \subcaption{Time}
        \includegraphics[trim=20 0 40 0,clip,width=0.75\columnwidth]{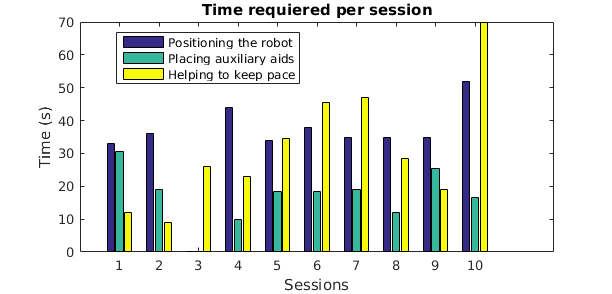}  
        \label{subfig:HelpTime}
    \end{subfigure}

    \begin{subfigure}[t]{0.85\textwidth}
        \centering
        \subcaption{Occurrences}
        \includegraphics[trim=20 0 40 0,clip,width=0.75\columnwidth]{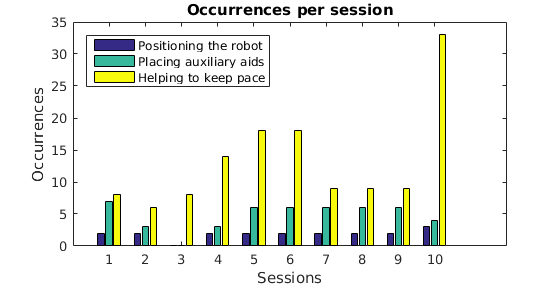}  
        \label{subfig:HelpOccu} 
    \end{subfigure}

\caption{Time required and number of occurrences per session for human-assisted actions.}
\label{fig:HelpCost}
\end{figure}

Figure \ref{subfig:HelpTime} provides a coarse-level analysis of time-costs associated with providing the
SAR assistance over ten patient sessions in Phase 2. Figure \ref{subfig:HelpOccu} shows the corresponding number of occurrences of each 
activity, for each session.  It can be seen that assisting the robot to keep pace (via head touch) required less time to perform, 
but occurred at significantly higher frequency than other human-assisted actions, scaling roughly with the number of activities to perform. 
While required often, Keeping Pace actions appeared to complement the general desire of patients to interact with the robot. 
Indeed, if close enough to the robot, and able, patients performed the action themselves. 
Therapist feedback indicated that allowing patients to deliver assistance to NAO also appeared to increase their activity and engagement during the session.  

Positioning the Robot and Placing Auxiliary Aids occurred less frequently than Keeping Pace actions, but as expected, required more session time to perform.  
However, therapists expressed no concern with this time cost (less than one minute), and thus we consider the SAR's human-assistance needs to be within  an acceptable  limit. 
Notably, however,  the exercise programs observed in the current study involve a relatively low number of human-assisted exercises.
We note that other rehabilitation programs may include a more diverse range of exercises that may require more carer assistance.

Physiotherapists participated in the study without any prior training, apart from being told that the SAR would ask for help from time-to-time. 
Therapists expressed willingness to provide assistance, and in general, demonstrated competence in 
handling the robot when required.
A notable issue that was observed in sessions was the therapist attempting to perform tasks for the robot that it was capable of itself.  In particular, laying the robot on its back for exercises.  Therapists were not explicitly told  
the SAR was capable of this itself, and thus understandably intervened.  Improvements to the SAR's instructions during
sessions, and more explicit statements of the SAR's 
capabilities during training should address this. 
In post session interviews, no concerns were expressed about the impact of the assistance they were required to provide.

\subsection{Design Process Evaluation}
\label{subsec:DesingProcessEvaluation}
The SAR has engaged with over 40 unique patients across both phases of development.  
Within 23 months, we have progressed from exploration activities during informal visits to a base-level 
stand-alone therapeutic aid for rehabilitation, deployed in weekly clinical sessions.  
Phase 1 was necessarily unstructured, employing in-situ \emph{Wizard-of-Oz} operation with therapists, patients and parents.
This is appropriate for busy clinical settings, but could be complimented by  formal requirements elicitation after a period of familiarisation.

Regular frequent in-situ engagement with clinical stakeholders has been key to establishing trust and rapport.  
During Phase 1, therapist attitudes evolved from curious and unconvinced at the beginning, to increasingly interested and engaged in the SAR's development,  and the design process.  
The design team now incorporates  technical, physiotherapy, cerebral palsy and psychology expertise.   
We argue that this in-situ design process has been essential to the establishment of the SAR as a legitimate and  viable therapeutic aid, which in turn has established
clinical advocates for the SAR.  This has been crucial to the recruitment of patients to participate in Phase 2 testing, and to
the long term support of the project by the rehabilitation clinic.

Phase 1 established researchers' relationship with clinical staff and clinical concepts. 
The identification of a set of exercises the robot was able to perform, and the clinical knowledge of a group of patients that commonly are prescribed those exercises was key. 
Defining the target patient population and associated exercise set in consultation with therapists in Phase 1 allowed therapists
to engage more directly with the design process by identifying appropriate patients to focus on, and to recruit for Phase 2 testing. 
Notably, in more recent Phase 2 testing the patient population has broadened to a larger population of children in rehabilitation,
suggesting the early focus on one patient cohort has not limited the scalability of the system to other patient groups.
 
We argue that the design of SARs for other health care applications may benefit from a similar design process of initial in-situ
 exploration and stake-holder relationship building, leading then to  
the focussed development of a viable prototype for feasibility and technical capacity testing in Phase 2.
We further advocate for a focus on discreet goals for the system, which in our experience allowed therapists to engage more readily with the
process.  Early Phase 1 attempts to present and demonstrate the general capabilities of the NAO system to therapists produced few 
outcomes, with no clear link to its practical implementation and therapeutic value.

The design process has provided therapists with direct access to the SAR system, allowing both hands-on experience 
manipulating robot limbs, but also with the software interface.  While in general health professionals  do not
have the time (and perhaps interest) in this level of access, our experience has been that physiotherapists generally take up the opportunity, when offered, to explore the SAR's capabilities.  
This was observed to increase familiarity with the SAR's capabilities (and limitations), but more
importantly, provided an entry point for care-givers to directly contribute to      
the requirements analysis and design of the SAR.
Whether the level of engagement we experienced is specific to physiotherapists, or to the 
particular clinic is unclear.  We argue, however, that providing frequent opportunities for stakeholders to engage with such
novel and unfamiliar technology promotes  transparency in the design process, and a sense of ownership of the deployed system.
This is a crucial feature
of any design process that seeks to deploy SAR's in a health care setting, where preconceptions
and a lack of familiarity and trust of the technology (and the design process) risks impeding confidence and acceptance.

Certain limitations should be considered when designing in-situ:
regular on-sight visitation requires large time investment of a small, dedicated technical development team.
Our approach promotes design and integration of an SAR into clinical practise but is not conducive to technical innovation by a small development team.
Parallel lab-based development could be informed by, and feed into Phase 2 prototype testing.
Stakeholders' expectations must also be managed. While in-situ development promotes design transparency, it also exposes delays and
system failures directly to end-users.  It is thus important to establish a common understanding of the constraints and limitations on both the system, and the development cycle.   

In-situ design in a health care setting must carefully manage all the above considerations within the
context of a highly demanding and busy clinical environment.   Technical developers must always concede to the needs of patients and therapists,
which may often mean little progress is made in an individual session.  High frequency visitation can mitigate this, increasing opportunities for
engagement with health care professionals, as well as their familiarity  and acceptance of the technical development team.

\section{Conclusion}
\label{sec:Conclusions}
We have presented our in-situ  design process  for the 
development of a socially assistive robot for paediatric rehabilitation.
Our two-phase process of exploration and development, embedded in the busy 
rehabilitation clinic of Melbourne's Royal Children's Hospital, has adapted a general purpose
off-the-shelf social robot, NAO, as a stand-alone therapeutic aid deployed and leading weekly rehabilitation sessions with patients.

We have listed a set of roles and requirements for our system, derived from an initial exploratory phase 
in order to develop our first prototype. We have explained the design considerations
in the current iterative development phase to satisfy the roles and requirements. 

A deliberately conservative system has been deployed.
While limited in capabilities, NAO's fast-tracked deployment as a 
robust minimalist system is providing crucial patient engagement
experience, and insights into what is required for ongoing clinical
deployment, and in particular, a formal clinical evaluation of its therapeutic benefits.    We argue that this approach has lead 
to a system that not only meets minimum operational and therapeutic requirements 
for clinical deployment, but also has clearly established priorities for further development as we prepare for  formal clinical trials of the SAR
for paediatric rehabilitation.  Such outcomes offer insights  to SAR design and development for other health care applications, particularly in busy clinic/hospital settings.



\begin{acks}
We gratefully acknowledge the physiotherapists, patients and parents involved in this study.
Project funding: Traffic Accident Commission (TAC) grant for Phase 1; 
and Data61, CSIRO student scholarship for Phase 2.
We also acknowledge The Brainary for their general support.
\end{acks}

\bibliographystyle{ACM-Reference-Format}
\bibliography{THRI2017}